\documentclass[twocolumn]{aastex701}

\usepackage{amsmath}
\usepackage{amsfonts}
\usepackage{amssymb}
\usepackage{upgreek}
\usepackage{bbold}
\usepackage{comment}

\usepackage{longtable}
\usepackage{array}

\newcommand\clearrow{\global\let\rowmac\relax}
\clearrow
\usepackage{supertabular}
\usepackage{tabularx}
\def\Rs{\hbox{R$_{\odot}$}} 
\def\kms{\hbox{km$\;$s$^{-1}$}} 

\begin{document}

\title{The near-Sun Heliospheric Current Sheet, fluid and kinetic properties}

 \author[orcid=0000-0001-6308-1715]{Naïs Fargette}
 \affiliation{Institut de Recherche en Astrophysique et Planétologie, CNES, CNRS, 31400 Toulouse, France}
 \affiliation{The Blackett Laboratory, Department of Physics, Imperial College, SW72AZ London, UK}
 \email{nfargette/at/irap.omp.eu} 

 \author[orcid=0000-0003-4733-8319 ]{Jonathan P. Eastwood}
 \affiliation{The Blackett Laboratory, Department of Physics, Imperial College, SW72AZ London, UK}
 \email{jonathan.eastwood/at/imperial.ac.uk} 

 \author[orcid=0000-0002-6924-9408]{Tai D. Phan}
 \affiliation{Space Sciences Laboratory, University of California, Berkeley, CA 94720, USA}
 \email{taiphan/at/berkeley.edu} 

 \author[orcid=0000-0002-6276-7771]{Lorenzo Matteini}
 \affiliation{The Blackett Laboratory, Department of Physics, Imperial College, SW72AZ London, UK}
 \email{l.matteini/at/imperial.ac.uk} 

 \author[orcid=0000-0002-7419-0527]{Luca Franci}
 \affiliation{
     School of Engineering, Physics and Mathematics, Northumbria University, Newcastle upon Tyne, NE1 8ST, UK}
 \affiliation{
     National Institute for Astrophysics (INAF)—Institute for Space Astrophysics and Planetology (IAPS), Rome, Italy}
 \affiliation{The Blackett Laboratory, Department of Physics, Imperial College, SW72AZ London, UK}
 \email{luca.franci/at/northumbria.ac.uk}

\begin{abstract}
The heliospheric current sheet (HCS) is an important large-scale structure of the heliosphere, and, for the first time, the Parker Solar Probe (PSP) mission enables us to study its properties statistically close to the Sun.
We visually identify the 39 HCS crossings measured by PSP below 50~\Rs~during encounters 6 to 21, and investigate the occurrence and properties of magnetic reconnection, the behavior of the spectral properties of the turbulent energy cascade, and the occurrence of kinetic instabilities at the HCS.
We find that 82\% of HCS crossings present signatures of reconnection jets, showing that the HCS is continuously reconnecting close to the Sun. 
The proportion of inward/outward jets depends on heliocentric distance, and the main HCS reconnection X-line has a higher probability of being located close to the Alfvén surface. 
We also observe a radial asymmetry in jet acceleration, where inward jets do not reach the local Alfvén speed, contrary to outward jets. 
We find that turbulence levels are enhanced in the ion kinetic range, consistent with the triggering of an inverse cascade by magnetic reconnection.
Finally, we highlight the ubiquity of magnetic hole trains in the high $\beta$ environment of the HCS, showing that the mirror mode instability plays a key role in regulating the ion temperature anisotropy in HCS reconnection.
Our findings shed new light on the properties of magnetic reconnection in the high $\beta$ plasma environment of the HCS, its interplay with the turbulent cascade and the role of the mirror mode instability.
\end{abstract}




\section{Introduction} \label{sec: 1_intro}
\begin{figure*}[t]
    \centering
    \includegraphics[width=\textwidth]{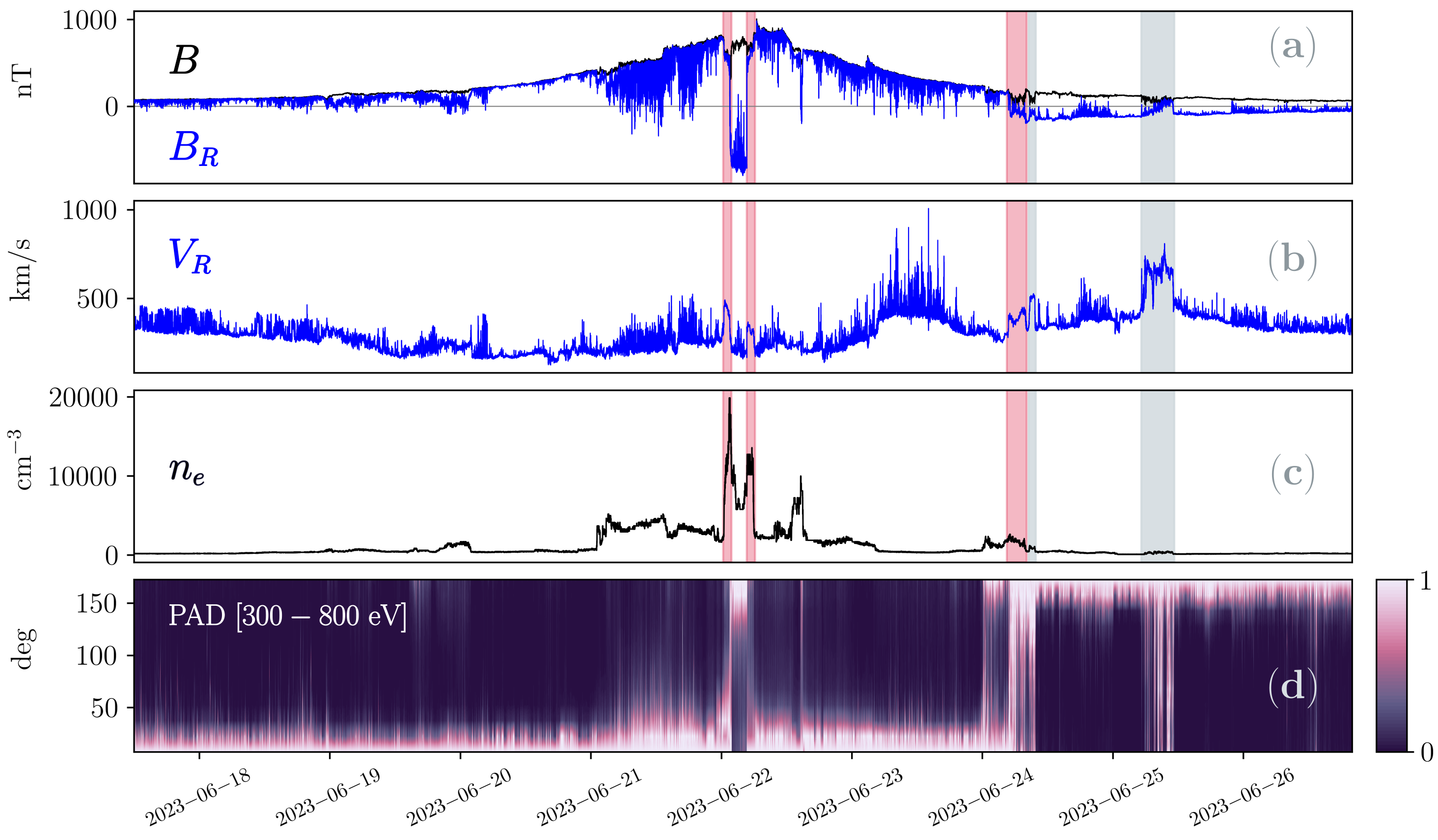}
    \caption{Identification of HCS full and partial crossings during E16, while PSP was located below 50~$R_{\odot}.$ From top to bottom, panels show (a) the magnetic field's amplitude $B$ and radial component $B_R$, (b) the radial solar wind velocity $V_R$, (c) the QTN electron density $n_e$ and (d) the PAD for suprathermal electrons (300-800~eV) normalized to its maximum value for each timestamp.
    The time intervals of HCS crossings are indicated by red shading, while partial crossings are shaded in gray.
    }
    \label{fig: E16}
\end{figure*}

The Heliospheric Current Sheet (HCS) is the structure separating the magnetic hemispheres of the Sun, the boundary between open solar magnetic field lines of opposite direction.
It is an important feature of our solar system that extends throughout the heliosphere, making it the largest current sheet observed \textit{in situ}.
Its configuration of antiparallel magnetic field lines makes the HCS a natural place to look for and study magnetic reconnection, in a system forced by the large-scale topology of the Sun.

Magnetic reconnection, a fundamental process in astrophysical systems, converts magnetic energy into kinetic and thermal energy, and enables large-scale remodeling of the magnetic field topology of astrophysical objects. 
It is at the crossroads of several fields of physics, playing a part in star-planet relations \citep[e.g.,][]{Gershman2024}, the dynamics of high-energy systems \citep[e.g.,][]{Guo_2024} and laboratory plasma physics \citep[e.g.,][]{Yamada_2022}.
In a non-colliding plasma, at first order, the triggering of magnetic reconnection is thought to depend on the thickness of the discontinuity \citep{Sanny_1994}, the magnetic shear at its boundaries and the ratio between thermal pressure and magnetic pressure $\beta = n k_B T \left(\dfrac{B^2}{2 \mu_0}\right) ^{-1}$ \citep{Swisdak_2003,Swisdak_2010,Phan_2010}, where $n$ and $T$ are respectively the plasma density and temperature, $B$ is the magnetic field amplitude, $k_B$ is the Boltzmann constant, and $\mu_0$ is the permeability of free space. 
However, the recent observations made by the Parker Solar Probe (PSP) mission have called into question the importance of these different factors.

Until recently, most available measurements of the HCS were made at the Earth orbit, and the properties of the current sheet at 1~AU have been well established \citep{Smith_2001}.
In particular, magnetic reconnection is rarely observed at 1~AU, and only a handful of cases have been reported \citep{Gosling_2005, Gosling_2006, Gosling_2007, Lavraud_2009}. 
By contrast, the first orbits of PSP with the Sun have unveiled that, in the inner heliosphere, magnetic reconnection signatures are frequently observed at HCS crossings \citep{Lavraud_2020, Szabo_2020, Phan_2021} despite a thickness theoretically inconsistent with magnetic reconnection onset \citep{Phan_2021}.
These past studies have been case studies, and a systematic study of HCS crossings is lacking. 
In this paper, we statistically study the occurrence and properties of HCS reconnection observed by PSP, using the complete set of publicly available PSP data (section \ref{sec: 4_Rx}).

As magnetic reconnection is found to be ubiquitous in the near-Sun HCS (see section \ref{sec: 4_Rx}), the latter is then a particularly interesting location to study the interplay between solar wind turbulence and magnetic reconnection. 
The power spectrum of the solar wind magnetic field typically exhibits a spectral break at ion kinetic scales, marking a separation between an inertial and a kinetic cascade \citep[see e.g.,][and references therein]{Bruno_Carbone_2013}.
The spectral break frequency decreases with heliocentric distance \citep{Bruno_Carbone_2013, Telloni_2015, Chen_2020, Duan_2020}. 
The impact of the HCS on the turbulent cascade near the Sun was studied by \citet{Chen_2021} during the 4$^{th}$ orbit of PSP. They showed that the properties of the turbulent cascade were significantly different between the near-HCS streamer belt plasma and the coronal hole solar wind, including a lower amplitude and a steeper magnetic field spectrum in the inertial range. In this work, we statistically study the spectral properties of the turbulent cascade within and around the HCS observed by PSP in section \ref{sec: 5_turbul}.

Finally, we report in section \ref{sec: 6_MH_train} on the ubiquity of magnetic hole (MH) trains observed inside HCS crossings close to the Sun. 
MH trains manifest as deep dips in the magnetic field amplitude $B$,  that occur in rapid succession. 
Usually thought to be associated with the mirror mode instability, MH trains in the solar wind are also referred to as "mirror mode storms" in the literature \citep{Russell_2009, EnriquezRivera_2013J, Dimmock_2022}.
Mirror mode storms, however, can also manifest as trains of peaks in $B$, but we only observe MH trains in HCS crossings. 
While mirror mode activity has been widely predicted, observed, and studied in planetary magnetosheath \citep{Chandrasekhar1958, Hasegawa1969, Tsurutani1982, Volwerk_2008, Soucek_2008, Genot_2009}, solar wind observations of mirror mode storms are rarer, usually associated with stream interaction regions or shocks \citep{Russell_2009, EnriquezRivera_2013J, Dimmock_2022}.
MH trains have also recently been reported within an HCS observed by the Solar Orbiter mission \citep{Dimmock_2022}.
We discuss the relation of these observed MHs to the mirror mode instability, and their role in plasma isotropization of the HCS. 

In this paper, we statistically investigate the structure of the HCS at different scales over 39~events in order to constrain the physics happening at this sector boundary in the inner heliosphere.
We present the data we use from PSP in section \ref{sec: 2_data}. 
In section \ref{sec: 3_large_scale}, we detail how we identified HCS crossings (\ref{subsec: 3.1_hcs_id}) and report on the HCS location and width (\ref{subsec: 3.2_hcs_loc_width}).
In section \ref{sec: 4_Rx}, we show that 82\% of HCS crossings present signatures of reconnection jets (\ref{subsec: 4_Rx_id}) and investigate the proportion of inward/outward jets as a function of radial distance, as well as their Alfvénicity and connectivity (\ref{subsec: 4.2}).
We then study how the spectral properties of the turbulent energy cascade behave within and around the HCS in section \ref{sec: 5_turbul}, highlighting increased levels of fluctuation below the ion scale in the HCS, as well as an attenuation of large-scale Alfvénic fluctuations in the close vicinity of the HCS.
In section \ref{sec: 6_MH_train}, we highlight the ubiquity of MH trains in the high $\beta$ environment of the HCS, hinting that the mirror mode instability plays a key role in regulating the plasma temperature in HCS reconnection.
We discuss the implications of our findings in section \ref{sec: discussion}.

\begin{figure*}[t]
    \centering
    \includegraphics[width=\textwidth]{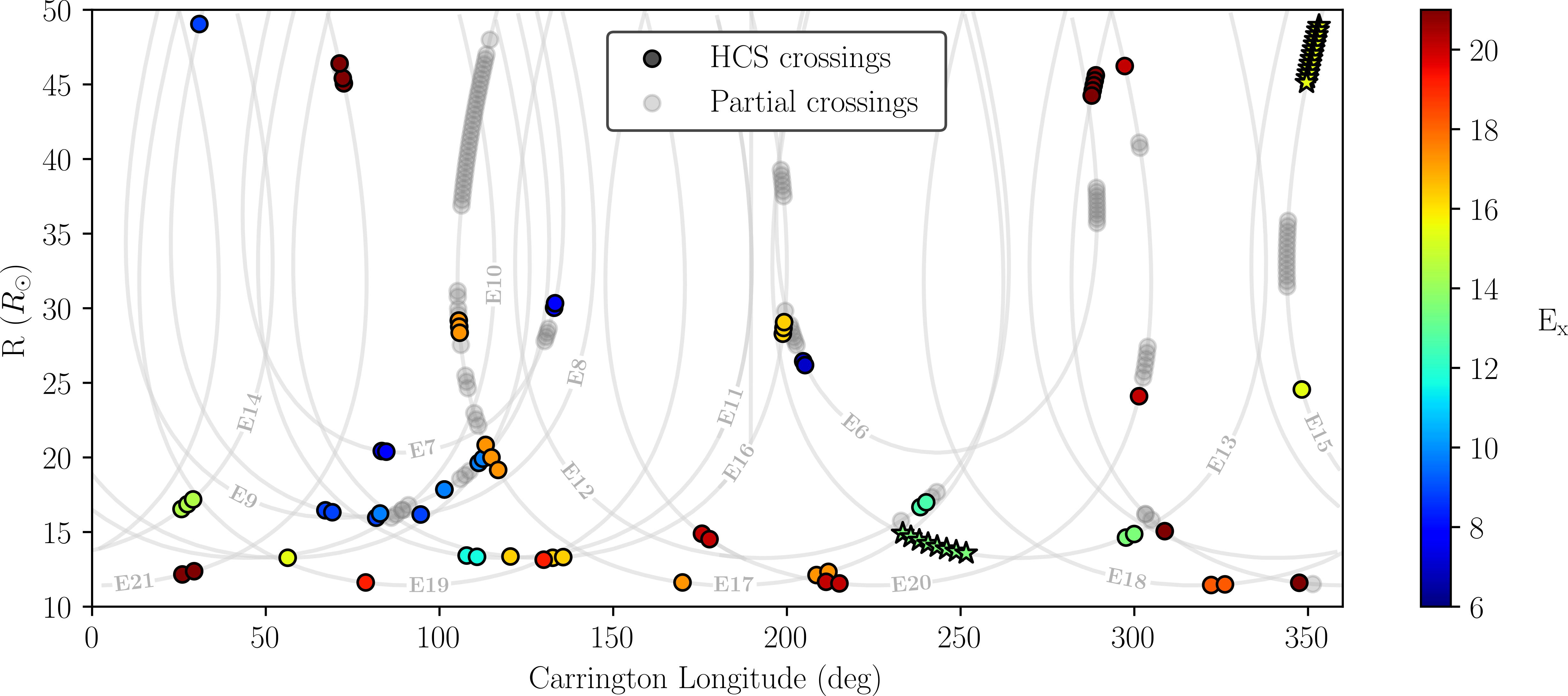}
    \caption{Radial distance and Carrington longitude of HCS crossings measured by PSP from E6 to E21. 
    We show PSP's orbits as grey lines and HCS crossings as full dots colored by encounter with a 1h time resolution. 
    We also show long partial crossings lasting more than 1h as light grey dots. CMEs from E13 and E15  -- associated with a change of polarity in the magnetic field -- are indicated with stars.
    }
    \label{fig: R_theta}
\end{figure*}

\section{Data} \label{sec: 2_data}

We investigate solar wind properties measured by the PSP mission.
We retrieve magnetic field and quasi-thermal noise (QTN) electron density measurements from the FIELDS instrument suite \citep{Bale_2016} and ion particle data from the Solar Probe ANalyzers \citep[SPANs][]{Whittlesey_2020, Livi_2021} of the Solar Wind Electrons Alphas and Protons (SWEAP) instrument suite \citep{Kasper_2016}. 
To ensure that the ion plasma moments from SPAN-ion are most likely to be accurate, i.e., that the solar wind is within the field of view (FOV) of the instrument, we restrict our analysis to periods where PSP was located below $50~\mathrm{R_{\odot}}$ throughout E6 to E21, where E$x$ stands for encounter number $x$ (E21 being the latest publicly available to this date).
To estimate the plasma density, we use electron QTN density when available and the SPAN ion density otherwise.
Data are shown in the $RTN$ frame of reference, with $\mathbf{R}$ (radial) being the Sun to spacecraft unit vector, $\mathbf{T}$  (tangential) the cross product between the Sun's spin axis and $\mathbf{R}$, and $\mathbf{N}$ (normal) completes the direct orthogonal frame.

\section{Large scale structure of the HCS} \label{sec: 3_large_scale}

\subsection{HCS  crossing identification}
\label{subsec: 3.1_hcs_id}
The first step of our study was to visually identify all the HCS crossings measured by PSP throughout E6 to E21. 
We based our identification on HCS classical \textit{in situ} signatures near the Sun: a reversal of the radial component of the magnetic field coincidental with a reversal of the strahl direction, i.e the pitch angle distribution (PAD) of suprathermal electrons direction, and an increase in the plasma density.
We excluded two time intervals where the change of magnetic polarity was directly a part of a coronal mass ejection (E13, E15).
These events have been further studied in the literature \citep[see e.g.][]{Romeo_2023, Dresing_2025} and are out of the scope of this analysis.
We set HCS boundaries based mainly on variations in PAD and magnetic field, with plasma density and velocity as additional indicators when necessary.
We checked that the solar wind was within the FOV of SPAN ion during our selected events, i.e that the peak of the velocity distribution function (VDF) was not observed on a boundary anode in azimuth. Only a handful of VDF did not meet this FOV criteria (0.3\% of the considered data, concentrated over two HCS crossings), and QTN data were available at those times. Therefore, it is unlikely that instrumental FOV effects impact our results.
In total, we identify 39~HCS crossings that are listed in Table~\ref{tab: app_HCS}.
HCS crossings are hereafter referred to using their associated number (column 1 of Table~\ref{tab: app_HCS}).

In Figure~\ref{fig: E16}, we show the result of our identification for E16.
Three main reversals of the PAD are identified as full HCS crossings (events \#18, \#19 \& \#20).
Two long partial crossings -- where PSP goes in and out of the HCS but remains in the same magnetic hemisphere -- are additionally highlighted in Figure~\ref{fig: E16}: one close to the full HCS crossing of June 24, and one isolated on June~25.
We include a list of partial crossings of the HCS in Table~\ref{tab: app_partials}, identified by a decrease in the magnetic field amplitude, an increase in density, and the presence in the PAD of either a strahl drop-out indicating field lines disconnected from the Sun, or bidirectional electrons indicating that both ends of the field line are connected to the Sun.
Since the focus of this paper is on full HCS crossings, this list of partial crossings may not be exhaustive.

In the rest of this work, we often need to define background quantities for the HCS, like density, magnetic field, solar wind speed, etc. 
We thus define, on each side of our events, time intervals that last 10\% of the HCS crossing duration.
We then define background values as the median of our different quantities computed over these disjoint time intervals that encompass the HCS crossing.

\subsection{HCS location and width}
\label{subsec: 3.2_hcs_loc_width}

In Figure~\ref{fig: R_theta}, we show the location of HCS detections in terms of radial distance $R$ and Carrington longitude. 
We detect one or several HCS crossings in all encounters except for E10, where only a partial crossing of the HCS was measured. 
They span all longitudes of the Sun, and detections are concentrated at lower radial distances consistently with the orbital bias of PSP. 
We see that no HCS crossing was measured between 31 and 44~\Rs, and these periods correspond to the co-rotation of PSP with the Sun for most encounters. 
Long partial crossings, however, are observed during co-rotation in some orbits (E6, E15, E16, E17, E20).
The two CME - HCS crossings  (E13, E15) are indicated with stars in Figure~\ref{fig: R_theta}.
The HCS longitude seems remarkably stable across close-by encounters, as highlighted by points of similar colors observed at similar longitudes for different orbits.
For instance, PSP crosses the HCS around 80-90$^{\circ}$ for E7, E8, E9, around 110-140$^{\circ}$ for E16, 17, 19, at both 175$^{\circ}$ and 210$^{\circ}$ for E17, E21, and around 290-325$^{\circ}$ for E18, E20, E21.

\begin{figure}[t]
    \centering    \includegraphics[width=0.49\textwidth]{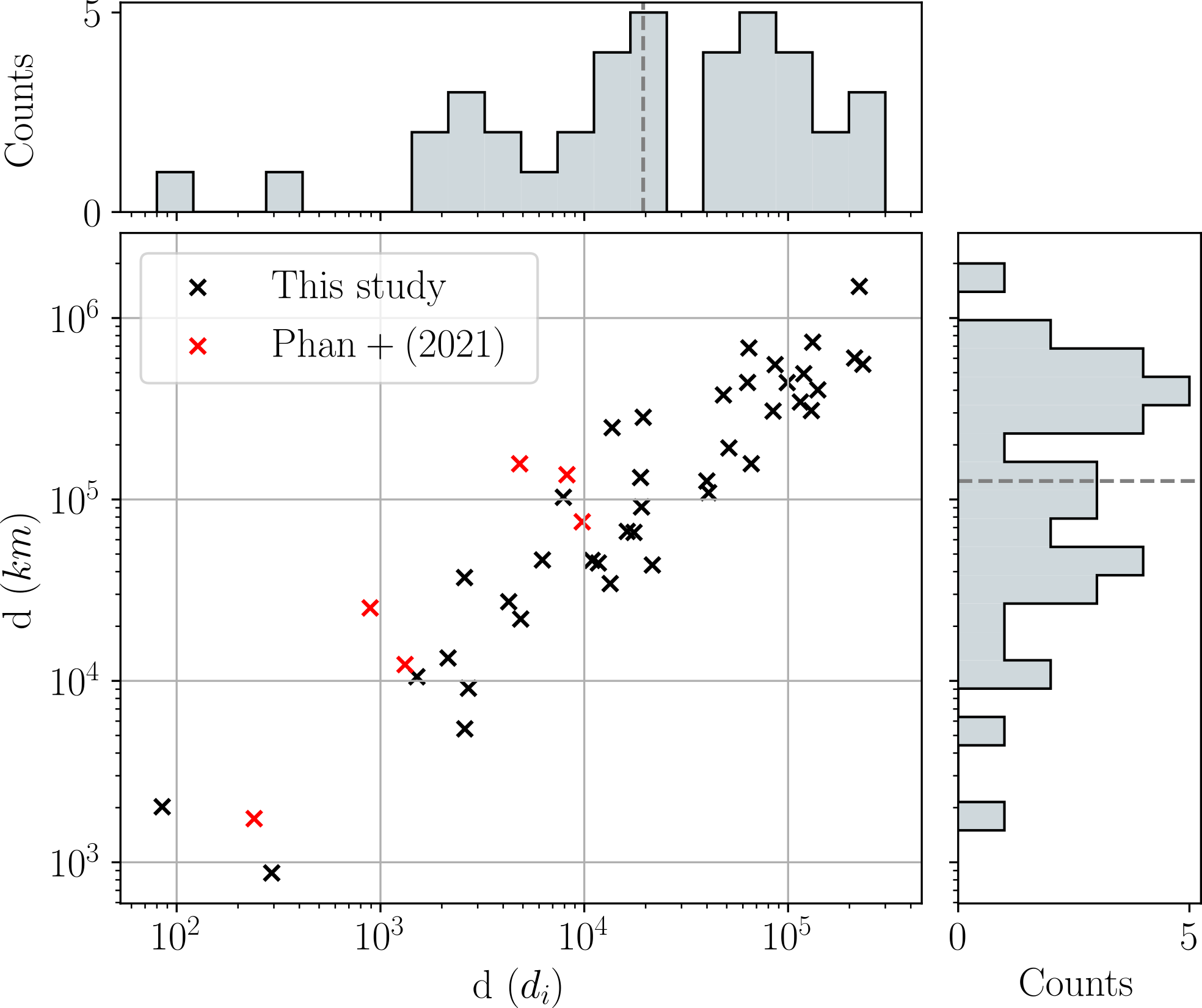}
    \caption{Width of HCS crossings, both in km and in ion inertial length ($d_i$) units. 
    Black crosses represent events from Table~\ref{tab: app_HCS} and red crosses are events from \citet{Phan_2021}.
    Top and side panels show the width distribution in $d_i$ and km, with median values indicated as a dashed line.
    }
    \label{fig: width}
\end{figure}

The HCS crossings we detect have a median (mean) duration of 25 (54) minutes, with values spanning from 35s (event \#36) to 4h52 (event \#34).
The longest events are usually located close to co-rotation periods, they correspond to PSP skimming the HCS.
To estimate the HCS width, we compute the vector normal to the HCS, $\mathbf{n}$, through a minimum variance analysis \citep[MVA,][]{Sonnerup_and_Cahill_1967}, on a time window that encompasses each HCS crossing with an additional 10\% duration on each side of the event. 
The $\mathbf{n}$ vector then points in the direction of minimum variance of the magnetic field.
We compute the width of the HCS as $d = |(\mathbf{V} - \mathbf{V_{PSP}}) \cdot \mathbf{n} | \delta t $, where $\mathbf{V_{PSP}}$ is the spacecraft velocity and $\delta t$ is the event duration.
We also compute the local background ion inertial length $d_i$ using the background density.

The ion inertial length $d_i$ varies from 2 to 23~km (with a mean value of $\overline{d_i}=6$~km). The ion inertial length linearly depends on heliocentric distance as expected, since $d_i \sim 1/\sqrt{n}$ and $n \sim 1/R^2 $ (not shown).
In Figure~\ref{fig: width}, we show the width distribution in both km and $d_i$ units.
These distributions stay consistent in order of magnitude when we vary the time window over which we perform the MVA, as well as when we use the hybrid-MVA method \citep{Gosling_and_Phan_2013}. 
The median HCS width is $3.6\times10^4~d_i$ ($1.6\times10^5$ km), which is higher than the width of the events reported in \citet{Phan_2021}, also included in Figure~\ref{fig: width}.
This higher width is mainly explained by the observation of longer events in this study ($\overline{\delta t} = 54$~min) compared to \citet{Phan_2021} ($\overline{\delta t} = 9.6$~min).
We further discuss the HCS location and the implication of such a large width in section \ref{subsec: 7.1_loc}, \ref{subsec: 7.2_reco}.

\section{Properties of HCS reconnection}
\label{sec: 4_Rx}

\subsection{Magnetic reconnection jet identification}
\label{subsec: 4_Rx_id}

\begin{figure}[t]
    \centering
    \includegraphics[width=0.48\textwidth]{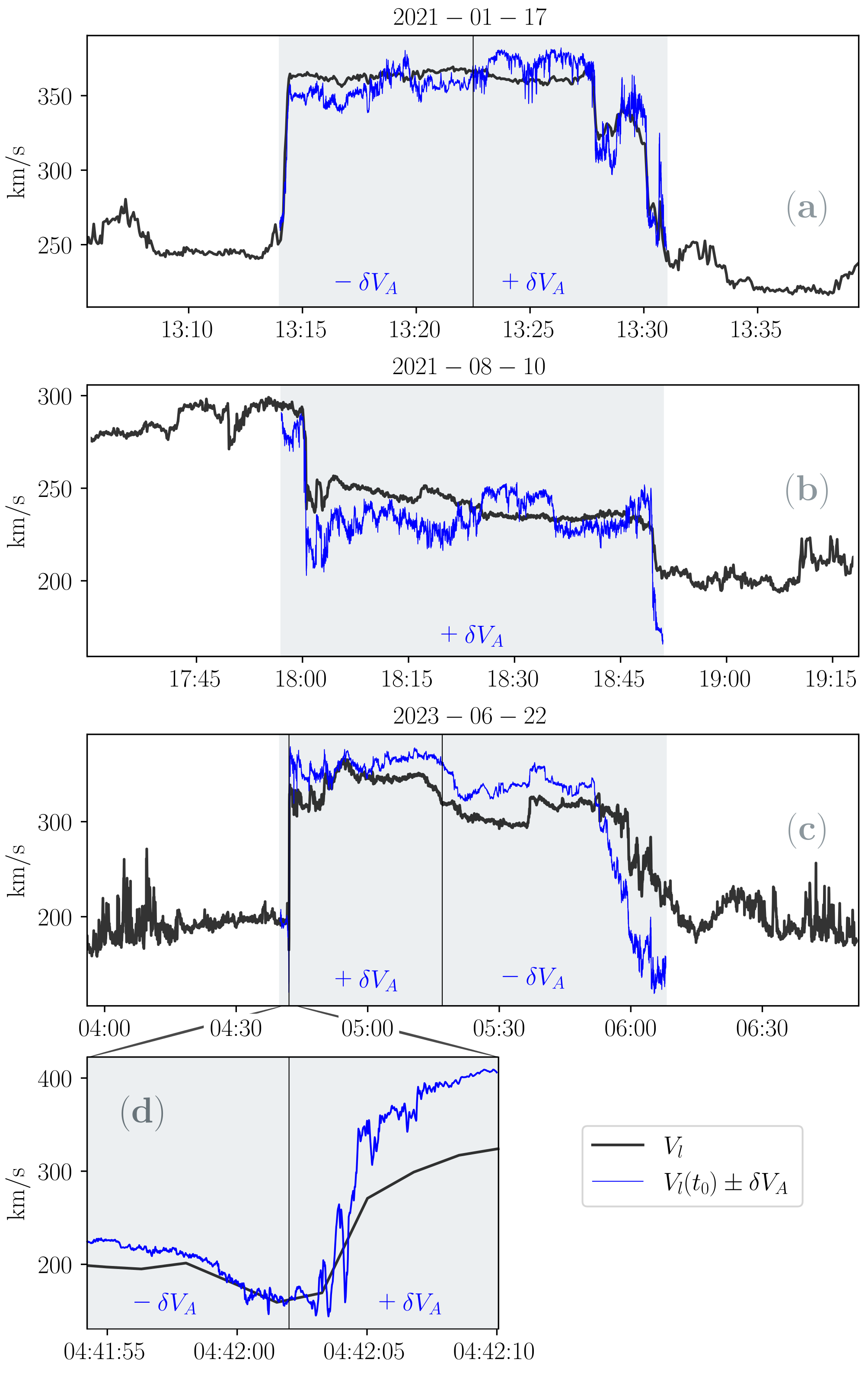}
    \caption{Examples of three representative HCS crossings. We show event \#11 where no ion jet is identified (a), event \#2 where a large-scale reconnection jet is observed (b), and event \#19 where both a large-scale jet (c) and a small-scale jet (d) are present.
    For each, we display in black the $l$ component of the solar wind velocity.
    We over-plot in blue $V_l(t_0) \pm \delta V_A$, where $t_0$ is an arbitrary time of the crossing. 
    Vertical black lines indicate a change in the correlation sign.
    }
    \label{fig: jets}
\end{figure}

\begin{figure*}[t]
    \centering    \includegraphics[width=1.\textwidth]{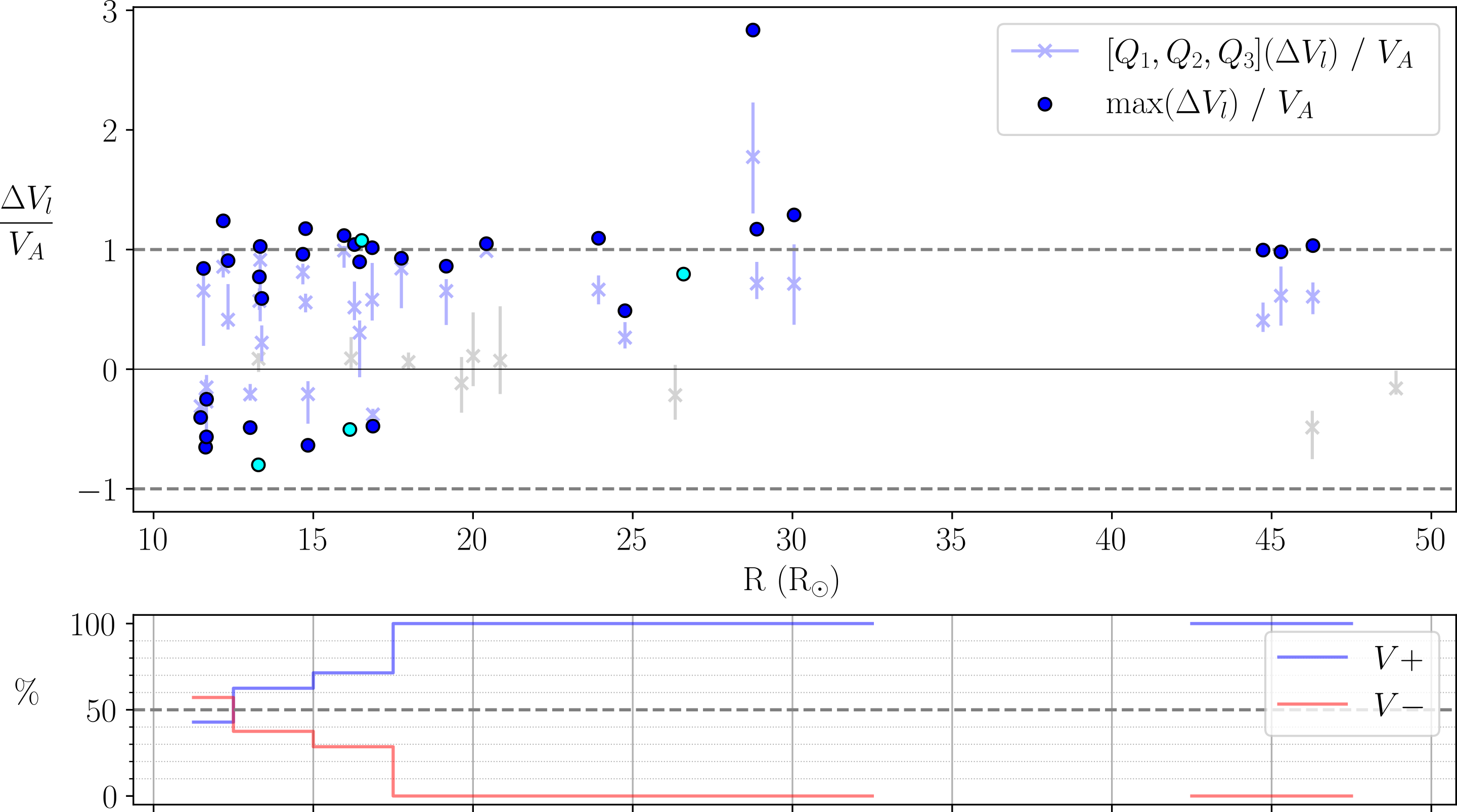}
    \caption{
    Properties of HCS reconnection jets. In the top panel, we show for each HCS the median velocity variation (i.e., the second quartile Q$_2$, crosses) along with the first and third quartile range $\mathcal{Q}_1 - \mathcal{Q}_3$ (vertical lines) of $\Delta V_l/V_{A}$ as a function of radial distance.
    Events with no large-scale jets are in gray, events with visible reconnection outflows are in blue.
    When reconnection was observed, we show the maximum velocity variation (full dots) normalized by $V_{A}$. 
    The small side jets of Table~\ref{tab: app_jets} are also included in lighter blue.
    In the bottom panel, we show the proportion of inward (red) and outward (blue) reconnection outflows observed by radial distance bins of 2.5~$R_{\odot}.$
    }
    \label{fig: dV_Va}
\end{figure*}
We then searched for signatures of magnetic reconnection jets in HCS crossings.
For each event, we rotated the data into its associated $lmn$ frame determined through an MVA of the magnetic field, where $l$ is the direction of maximum variance of the magnetic field.
The MVA method is adapted to the high shear configuration of the HCS. 
We checked for the presence of an ion jet in the $l$ direction coincidental with the magnetic field rotation \citep{Gosling_2005_JGR}.
We also checked that the correlation signs between the solar wind velocity $\mathbf{V}$ and the magnetic field $\mathbf{B}$ were opposite on each side of the jet. 
This correlation change is consistent with rotational discontinuities bounding the reconnection jet, with $\delta \mathbf{V} = \pm \delta \mathbf{V_A}$ at each boundary, where $\mathbf{V_A}=\mathbf{B}/\sqrt{\mu_0 \rho}$ is the Alfvèn speed, and $\rho$ is the proton mass density. 
During a reconnection jet, we also expect a plasma density increase, a drop in magnetic field amplitude, and PAD signatures consistent with a change in connectivity. 
We treated these signatures as additional hints of magnetic reconnection occurring, without being sufficient on their own.
Over the 39 HCS crossings studied, we find that 32 (82\%) include signatures of a magnetic reconnection jets, while 7 (18\%) present no ion jet signature that we could clearly identify (see Table~\ref{tab: app_HCS}).

In our database, some events are straightforward to analyze, while others require more careful treatment.
In Figure~\ref{fig: jets}, we show the $l$ component of the solar wind velocity and associated Alfvén speed variations for three representative HCS crossings.
In panel a, we display one of the seven HCS crossing where we did not detect reconnection signatures, and the solar wind velocity is positively correlated to the magnetic field throughout.
Then, for most events (28/39), the increase in ion velocity associated with reconnection coincides with the HCS structure. 
In panel b, we showcase an Alfvénic reconnection jet that is negatively correlated to the magnetic field at the leading HCS boundary, and positively correlated at its trailing edge.
Then, for 4/39 events, reconnection was detected but the HCS signatures and ion jet signatures do not fully coincide (events \#1, \#5, \#7 and \#19).
Magnetic field, PAD and ion properties delineate a large-scale structure that we identify as the HCS.
It may be associated with a reconnection-associated velocity increase (\#5, \#19) or present no overall acceleration (\#1, \#7).
In parallel, a small-scale ion jet is detected at $B_R$ reversal in these 4 events. For event \#19, showcased in Figure 4c and 4d, this leads to two reconnection jets being identified: one is large scale and associated with the full HCS crossing, the other lasts a few seconds and is located at the $B_R$ reversal. These 4 ion jets that take only a fraction of the HCS crossing at $B_R$ reversal are listed in Table~\ref{tab: app_jets}.
These HCS crossings are reminiscent of similar observations made at 1~AU \citep{Eriksson2022}, interpreted as a fracture of the HCS into many secondary reconnecting current sheets.
We note that smaller ion jets stemming from turbulence or flux rope interaction may also be present inside HCS crossings, akin to those reported by \citet{Phan_2024}.

\subsection{HCS outflow properties}
\label{subsec: 4.2}

Among the 34 reconnection jets we detected (30 listed in Table~\ref{tab: app_HCS} and 4 in Table~\ref{tab: app_jets}), 25 jets were directed outward and 9 jets were directed inward, i.e., towards the Sun. 
For these sunward cases, PSP was then located in between the Sun and the reconnection X-line, in a configuration similar to the one described in \citet{Phan_2024}.
In Figure~\ref{fig: dV_Va}, we see that all of the inward reconnection jets are observed between 11 and 17~\Rs.
While we only observe outward jets above 20~\Rs, the inward/outward proportion is 37.5/62.5 in the 12.5-15~\Rs~bin (8 events), and is inverted to 57/43\% below 12.5~\Rs~(7 events).

Another important feature of HCS reconnection is the jet speed in the HCS relative to the ambient solar wind flows.
To estimate it, we focus on the variation in $V_l$, the velocity change in the direction of maximum variation of the magnetic field, usually very close to the radial direction.
We first linearly detrend $V_l$ to remove the background component and obtain $\Delta V_l(t) = V_l(t) - V_{bg}(t)$, where $V_{bg}(t)$ is a linear interpolation between the median background velocities on each side of the HCS crossing (see appendix \ref{sec: app_vl} for more details).
We also compute the local hybrid Alfvén velocity \citep{Cassak_shay_2007},
\begin{equation}
V_{A}^{~2} = \dfrac{B_{l1}B_{l2}(B_{l1} + B_{l2})}{\mu_0 (\rho_1 B_{l1} + \rho_2 B_{l2})}
\end{equation}
where subscripts $_1, _2$ refer to background values computed before and after the event.
We show different relevant values of $\Delta V_l$ for each HCS crossing in Figure~\ref{fig: dV_Va}. 
\begin{figure}[b!]
    \centering    
    \includegraphics[width=.48\textwidth]{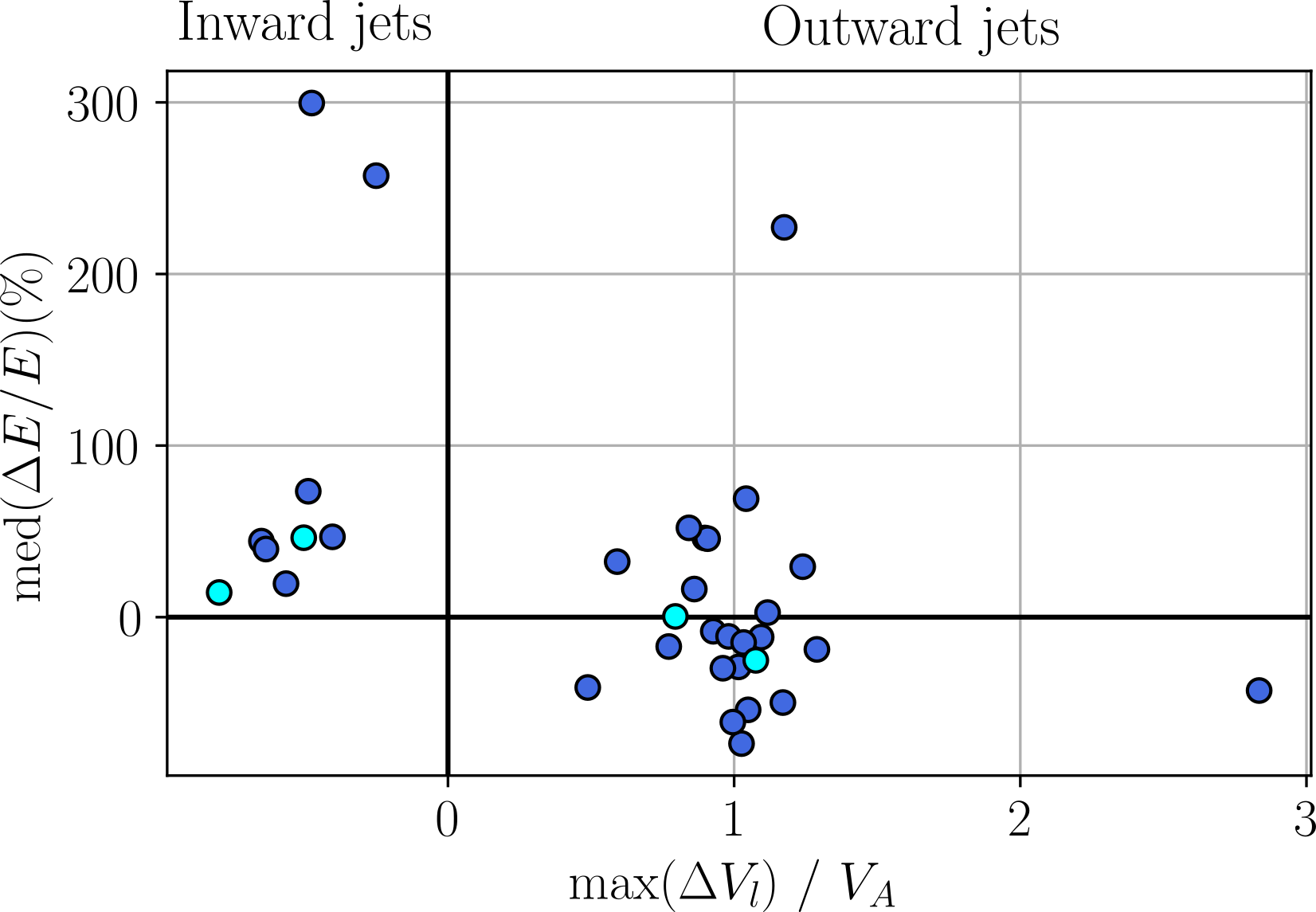}
    \caption{
    Median value of the relative variation of the energy flux $\Delta E / E$ for suprathermal electrons (300-800~eV) during HCS crossings, as a function of the maximum velocity variation (see appendix \ref{sec: app_vl} for computation details).
    Dark blue dots are reconnection jets from Table~\ref{tab: app_HCS}, while small jets from Table~\ref{tab: app_jets} are in light blue.
    Inward jets are on the left hand side ($\Delta V_l <0$) and outward jets are on the right hand side ($\Delta V_l >0$)
    }
    \label{fig: dE}
\end{figure}

What is striking in Figure~\ref{fig: dV_Va} is that outward reconnection jets statistically reach the local Alfvén speed, with $\langle \max(\Delta V_l) / V_A\rangle = 1.0$, where $\langle\cdot\rangle$ denotes the average over the considered events.
By contrast, the inward  jets are sub-Alfvénic and only reach $\langle\max(\Delta V_l) / V_A\rangle = -0.5$ on average.
A particular event (\#21) seems to reach almost twice the local Alfvén speed on average, with a maximum at $\Delta V_l / V_A = 2.9$. 
This HCS crossing is associated with the inbound of E17, where PSP was continuously skimming the HCS for several days. 
There, HCS velocity increases associated with nearby partial crossings are consistently displaying $\Delta V_l \sim$ 300~\kms, while the local Alfvén speed was no more than 200~\kms. 
Further investigation into the E17 inbound is required to understand this particular observation.

\begin{figure*}[t]
    \centering    \includegraphics[width=.9\textwidth]{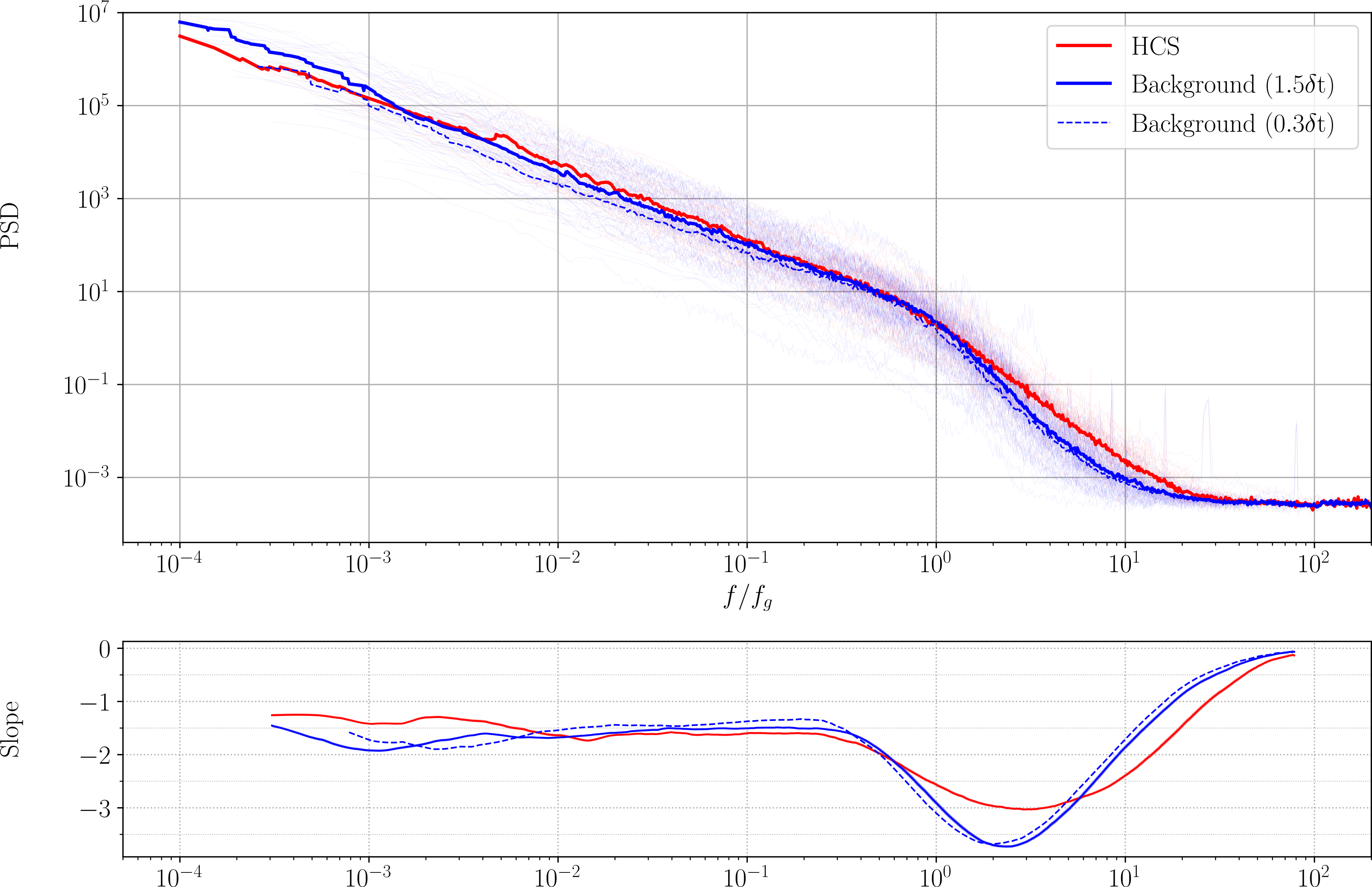}
    \caption{
    Power Spectral Density (PSD) of the magnetic field inside and outside HCS crossings. 
    Frequencies are normalized to the ion gyro-frequency $f_g$.
    For each event, we show in the top panel the magnetic field PSD inside the HCS crossing (thin red) as well as outside, using time windows lasting both 30\% (not shown) and 150\% of the crossing duration (thin blue).
    Thick curves are the median of all the individual PSDs thus computed.
    The thick dashed blue line represents the solar wind PSD near the HCS, while the thick solid blue line captures a more distant background.
    The thick red curve is the median PSD of the magnetic field inside HCS crossings.
    Local spectral slopes of the median PSD for each of these regions are shown in the bottom panel. 
    }
    \label{fig: PSD}
\end{figure*}
A feature shared by all inward reconnection jets is the presence of counter-streaming suprathermal electrons (Figure~\ref{fig: app_dV_dE}), as well as an increased energy flux of the strahl. 
In Figure~\ref{fig: dE}, we show the relative variation in energy flux for suprathermal electrons during reconnection outflows (see appendix \ref{sec: app_vl} for more details). 
Among 25 outward jets, 15 show a decrease in energy flux consistent with magnetic field lines being disconnected from the Sun, 8 present an increased $\Delta E / E$, and 2 have a constant energy flux ($\Delta E / E< 5\%$). 
One outward outflow shows a 227\% increase in $\Delta E / E$; it corresponds to E13 (event \#14), where the HCS was strongly influenced by the preceding CME \citep{Phan_2025}.
By contrast, all inward jets present an increased energy flux in the outflow, event \#15 and \#31 even reaching respectively a 300\% and 257\% increase.
This signature of increased strahl energy flux in inward jets, together with the presence of counter-streaming electrons (systematically observed, not shown), indicates that PSP is located on closed magnetic field lines connected on both ends to the Sun.
We further discuss the properties of HCS outflows in section \ref{subsec: 7.2_reco}.

\section{Spectral properties} \label{sec: 5_turbul}

We now investigate how the spectral properties of the turbulent cascade behave inside and around the HCS. 
In Figure~\ref{fig: PSD}, we show the power spectral density (PSD) of the magnetic field inside and outside HCS crossings, with frequencies normalized to $f_g$, the ion gyro-frequency.
First, we can clearly identify the spectral break around the ion gyro-frequency in the median spectra of both the background (blue) and the HCS (red).
Figure~\ref{fig: PSD} also shows that the PSD of the magnetic field is enhanced within the HCS compared to background fluctuation levels at frequencies higher than the ion gyro-frequency ($f/f_g$ = 1--20).
There, in the dissipation range, the slope of the HCS PSD is consistently equal to -3, while that of the background PSD briefly reaches -3.7 before hitting the noise floor of the instrument (showing as a flattening of the spectrum at higher frequencies). Our findings are consistent with simulation work showing that magnetic reconnection enhances turbulence levels locally, as discussed in section \ref{subsec: 7.3_turbulence}.

In the inertial range, the slope of the 30\% background, 150\% background, and HCS PSDs are quite close, respectively -1.4, -1.5, and -1.6 in the $10^{-2}$--$1$ frequency range.
It is interesting to note that background fluctuations present lower energy levels close to the HCS in the inertial range, as highlighted by the lower 30\%-background PSD (dashed blue) compared to the 150\%-background PSD (full blue) for $f/f_g<1$.
This means that large-scale fluctuations of the magnetic field are attenuated near the HCS. When browsing the data, we visually noticed that large Alfvénic deflections of the magnetic field, also referred to as magnetic switchback in the literature \citep[see e.g.,][]{Bale_2019, Raouafi_2023} seem to disappear right before and after HCS crossings. The magnetic field systematically becomes quiet and smooth around the HCS, which is visible by eye in Figure~\ref{fig: E16} and appears statistically in the average PSD of the magnetic field in Figure~\ref{fig: PSD}, where the spectrum close to the HCS contains less power compared to the spectrum further out.
We discuss a potential explanation linked to spacecraft connectivity in section \ref{subsec: 7.3_turbulence}.

\section{Magnetic hole trains in the HCS} \label{sec: 6_MH_train}

\begin{figure}[b!]
    \centering  \includegraphics[width=.49\textwidth]{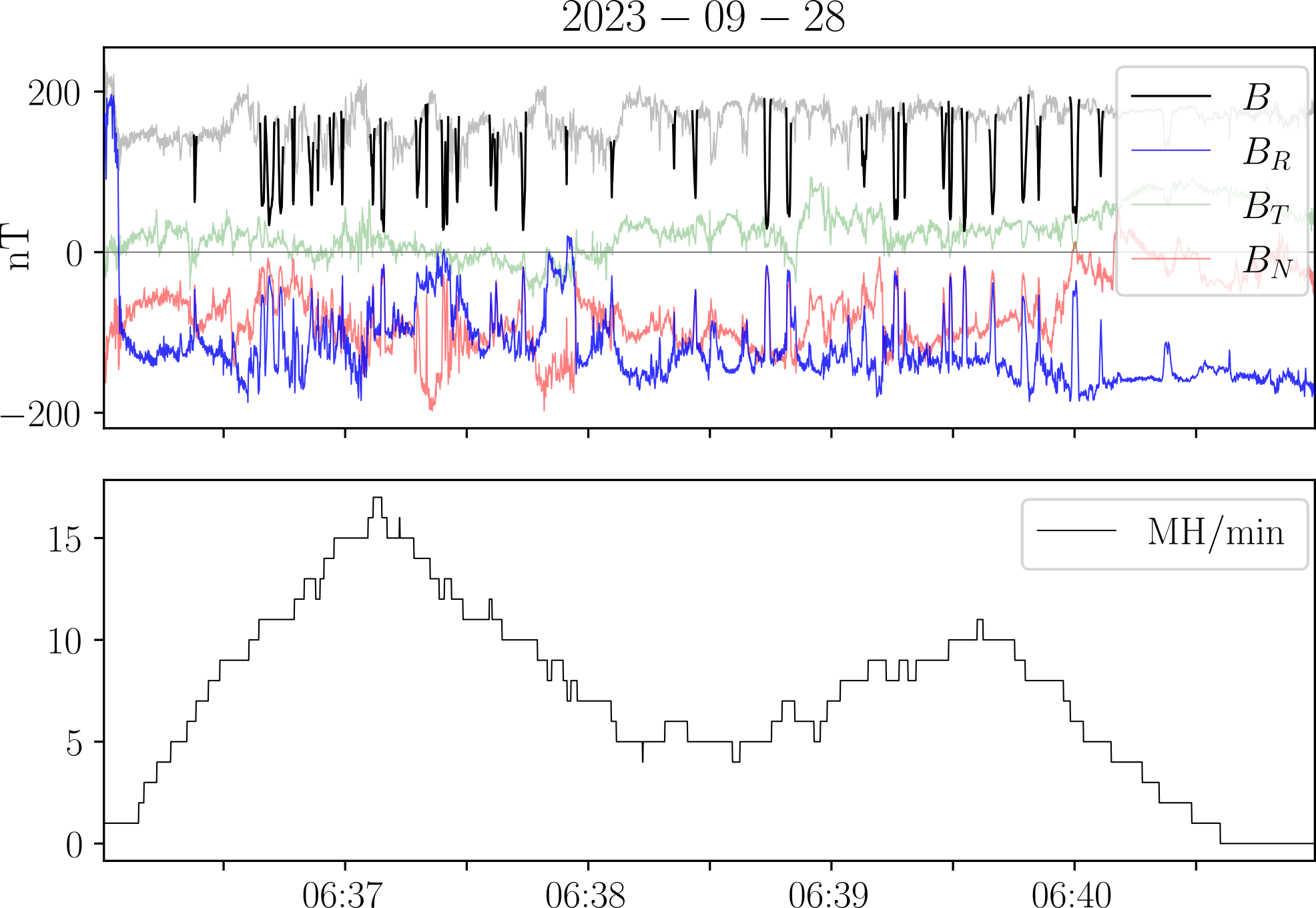}
    \caption{Illustration of a magnetic hole train (also called a mirror mode storm) during event \#26.
    Top panel shows the magnetic field with detected MH highlighted in black in the magnetic field amplitude.
    Bottom panel shows the MH occurrence rate.}
    \label{fig: mag_holes}
\end{figure}
Finally, we report on magnetic hole (MH) trains being observed almost systematically by PSP inside HCS crossings.
MH trains manifest as deep dips in the magnetic field amplitude that occur in rapid succession. 
For each HCS crossing, we automatically detect the presence of MHs using the algorithm described in appendix~\ref{sec: app_mh}. 
Dips in the magnetic field are labeled MHs if they reach a 35\% decrease compared to the background field \citep[similarly to ][]{Madanian_2020}.
In total, we detect 3423 MHs inside the 39 HCS crossings. By contrast, we only detect 175 MHs outside of HCS crossings, on time intervals lasting 50\% of $\delta t$ on each side of HCS crossings, meaning that the outside detection intervals have the same duration as the HCS crossings.

\begin{figure}[t]
    \centering  \includegraphics[width=.49\textwidth]{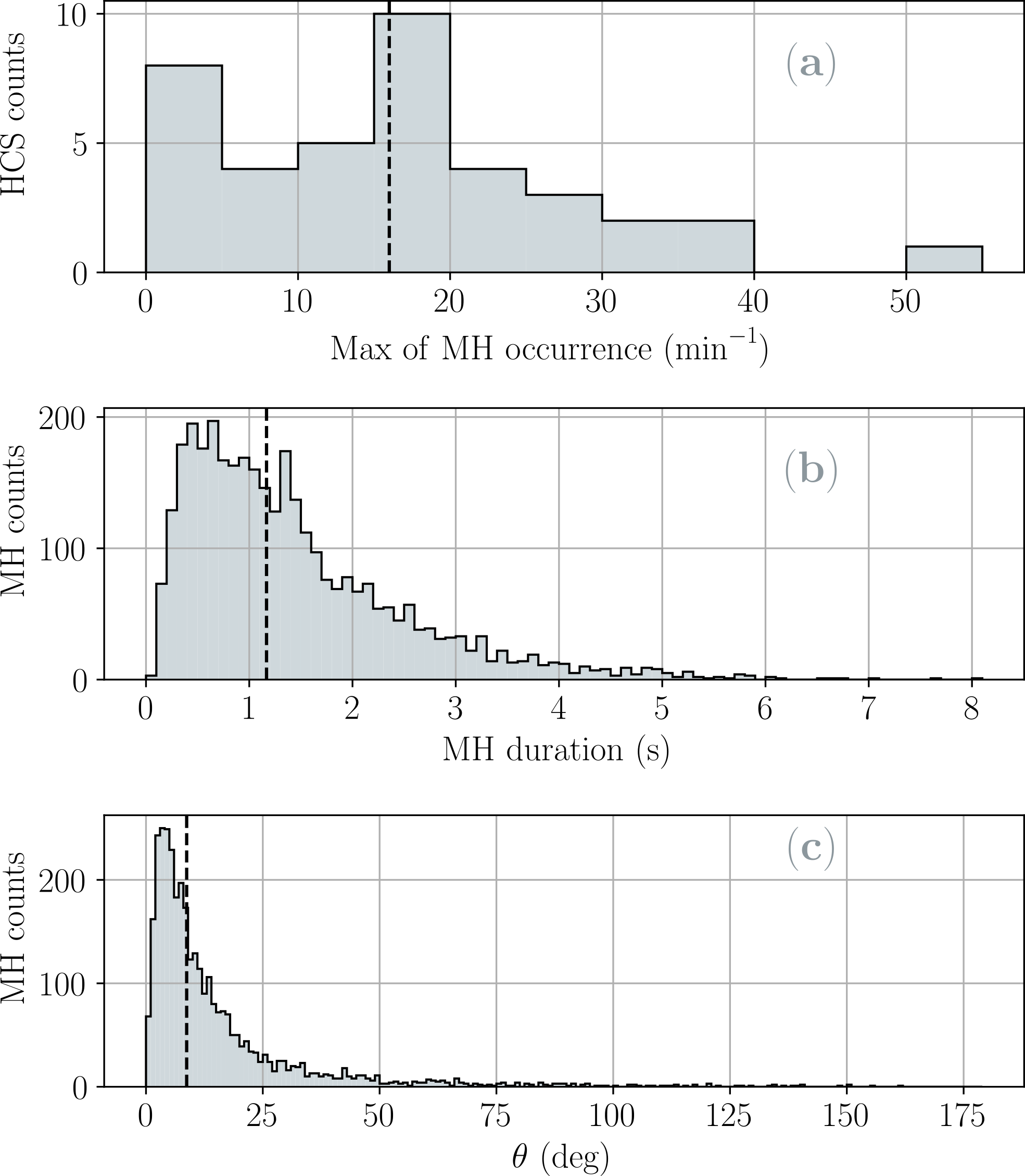}
    \caption{Properties of MH in HCS crossings. 
    In panel (a), we show the distribution of maximum occurrence of MHs reached during HCS crossings. In panel (b) and (c), we show respectively the distribution of MH duration and MH magnetic field rotation $\theta$.}
    \label{fig: MH_stats}
\end{figure}

In Figure~\ref{fig: mag_holes}, we show the detection result for a time interval of a few minutes, located within an HCS (event \#26), right after a sharp reversal of $B_R$.
The occurrence rate of MHs reaches 15 min$^{-1}$ during this time period.
We set an ad-hoc threshold of 5 MH/min to define a MH train, and we visually confirm or reject their presence for each crossing.
In Figure~\ref{fig: MH_stats}, we show the maximum occurrence rate reached in each HCS crossing (panel a).
Overall, MH trains are observed in 31 HCS crossings out of 39. 
The median maximum occurrence rate is 16~min$^{-1}$, with a maximum of 54~min$^{-1}$ observed during event \#37, where MH trains are particularly striking.
Among the 8 events without MH trains, 4 include isolated MHs with an occurrence rate lower than 5~min$^{-1}$ (\#7, \#29, \#32, \#36), and 4 show a complete absence of MH (\#4, \#15, \#19, \#31).

In Figure~\ref{fig: MH_stats}, we also display the duration (panel a) and $\theta$ (panel b) distribution of the 3423~MHs detected during the 39 HCS crossings, where $\theta$ is the angle of rotation of the magnetic field inside MHs (taken at the global minimum of the magnetic field, see appendix \ref{sec: app_mh}). 
The detected MHs have a median duration of 1.17~s, they are generally above the ion scale ($f_g \in [1-16]$~Hz).
They are linearly polarized, with a median rotation of 8.7$^{\circ}$ and with 90\% of the $\theta$ distribution below 34$^{\circ}$.
These properties are consistent with previously observed mirror mode storms, as discussed in section \ref{subsec: 7.4_MH}.

\section{Discussion} \label{sec: discussion}

The HCS is an important large-scale structure of the heliosphere, and, for the first time, the PSP mission enables us to statistically study its properties close to the Sun.
We identify and study the 39 full HCS crossings measured by PSP at heliocentric distances lower than 50$R_{\odot}$ throughout E6--E21.

\subsection{HCS location}
\label{subsec: 7.1_loc}
In terms of where the HCS is observed (Figure~\ref{fig: R_theta}), we find HCS  crossings at all Carrington longitudes, and its location is sometimes stable across several consecutive orbits. This stability is consistent with 1~AU results, where magnetic sectors are seen to be stable across Carrington rotations even during the rising phase of solar maximum \citep{Sheeley_Harvey_1981}. 
We observe HCS crossings between 11 and 49~$R_{\odot}$, but find a lack of HCS observation in the 31--44~\Rs range. 
This corresponds to periods where PSP was in co-rotation with the Sun, lowering the probability of measuring a full HCS crossing. 
During one encounter (E17), PSP did hover above the HCS during most of its inbound approach to perigee, and long partial crossings of the HCS are then observed.
Future investigation of this encounter using multi-spacecraft analysis to reconstruct the HCS structure and orientation \citep[similarly to][]{Laker2021} could greatly improve our understanding of the HCS temporal dynamics.

\subsection{Magnetic reconnection outflows at the HCS}
\label{subsec: 7.2_reco}
We find that the HCS almost always presents signatures of magnetic reconnection jets, with reconnection jets identified in 82\% (32/39) of HCS crossings.
This is consistent with previous studies on early PSP encounters that noted this prevalence in a handful of cases \citep{Lavraud_2020, Szabo_2020, Phan_2020, Phan_2021}.
It is the first time that this occurrence rate of reconnection is quantified, taking into account PSP encounters beyond E4, and thus reaching lower radial distances. 
This prevalence of reconnection is observed despite the HCS being very thick, 10$^3$--10$^5~ d_i$ (Figure~\ref{fig: width}), which is wider than previously reported close to the Sun \citep{Phan_2021}. 
This width of $1.6 \times 10^5$~km on average is remarkably similar to 1~AU observations \citep{Liou_2021}, showing that the HCS thickness is not varying significantly with radial distance.
This thickness, much larger than the ion inertial length at which reconnection is theoretically triggered \citep{Sanny_1994, Swisdak_2003}, is then consistent with a picture of the near-Sun HCS being constituted mainly of bulging flux ropes that quickly expand with distance to the X-line \citep{Phan_2024}.
The widths we find are also consistent with the size of HCS blocs imaged by PSP \citep[0.5-3~\Rs, i.e. 10$^5$--10$^6$~km, ][]{Liewer_2024}.
This all fits with the picture of fast ideal tearing mode \citep{Pucci_2014} being triggered at the HCS, as simulated by \citet{Reville_2020}.

\begin{figure}[b]
    \centering  \includegraphics[width=.49\textwidth]{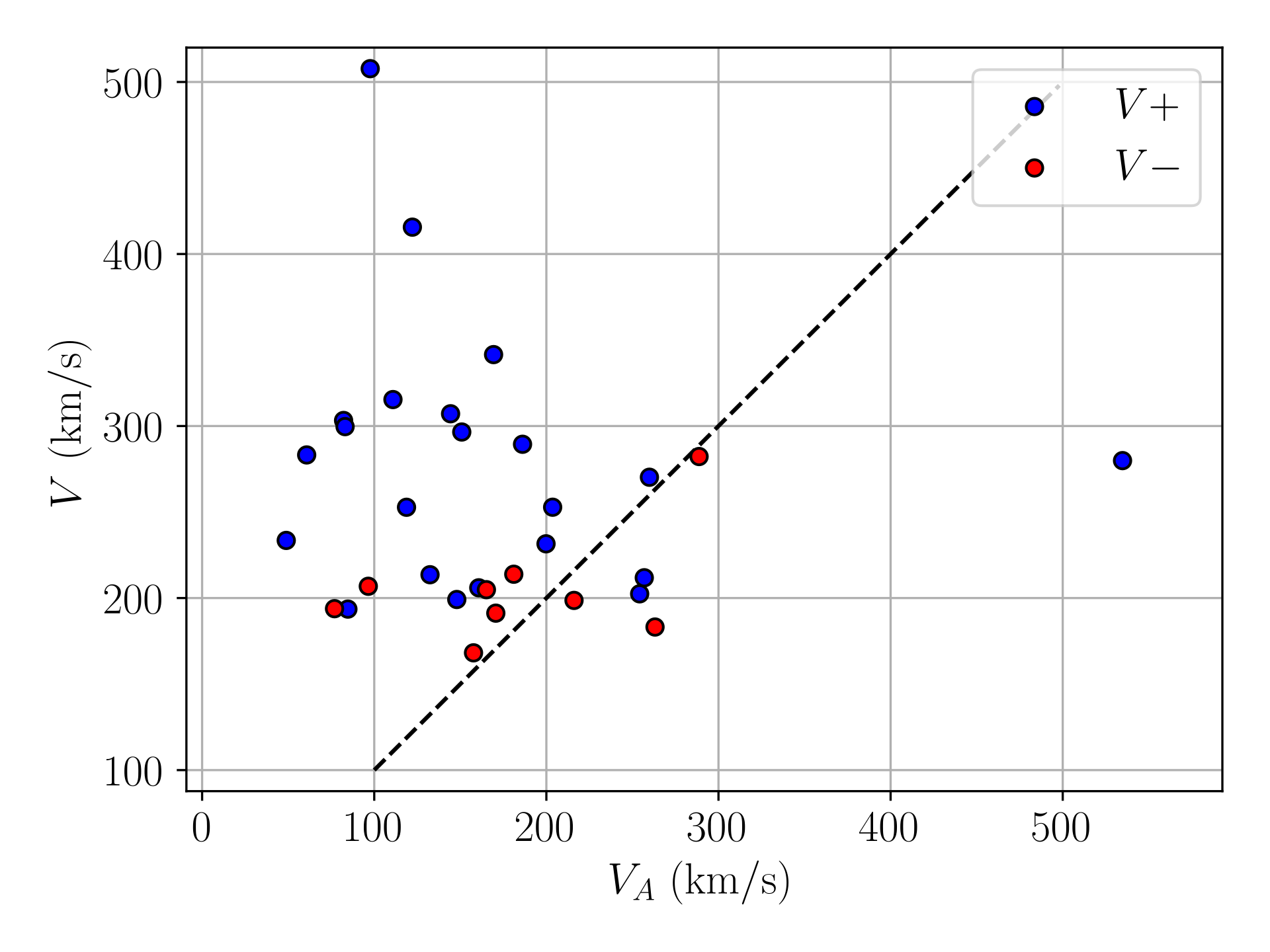}
    \caption{Background solar wind ion velocity as a function of background hybrid Alfvén speed for reconnecting HCS crossings. 
    The dashed line is $V = V_A$. 
    HCS crossings where a sunward reconnection jet was observed are in red.}
    \label{fig: V_Va}
\end{figure}

We further study the properties of HCS reconnection and report 9 inward jets and 25 outward jets.
We find that the proportion of inward/outward jets depends on the heliocentric distance (Figure~\ref{fig: dV_Va}), starting at a 100\% of outward jets above 17~\Rs, and decreasing towards a 50\% proportion around 12.5~\Rs.
We even see an inversion in the 10--12.5~\Rs bin, with more inward jets being observed, but this is most likely not significant.
While we only observe outward jets above 17~\Rs, another sunward jet associated with the HCS has been observed at 36~\Rs~during E4 \citep{Phan_2021}, and a few have been identified at 1~AU \citep{Gosling_2006, Lavraud_2009}.
As PSP probes closer distances to the Sun, we approach a 50\% partition of inward/outward jets, i.e., an equal probability of being located on either side of the X-line. 
This equipartition of inward/outward jets at low radial distance seems consistent with an active X-line preferentially located in the 10--17~\Rs region. 
In the fast tearing mode simulation of \citet{Reville_2020}, the primary reconnection developed in the 2.5--10~\Rs range before being advected in the solar wind.
We investigated the occurrence of sunward jets and show in Figure~\ref{fig: V_Va} that these sunward jets are actually more likely to be observed when the background solar wind is close to or lower than the background Alfvén speed.
However, some sunward jets are observed in the $V>V_A$ region, meaning that HCS reconnection is sometimes still active above the Alfvén surface.

In parallel, we observe that outward jets reach 100\% of the local Alfvén speed on average, while inward jets are only accelerated to 50\% of the local Alfvén speed (Figure~\ref{fig: dV_Va}). 
The peak of the ion VDF was within the FOV of SPAN for these sunward jets, which weakens a potential instrumental explanation of this asymmetry.
We also find increased suprathermal electron energy content for inward jets compared to outward jets (Figure~\ref{fig: dE}), which is consistent with a closed field line topology with both ends anchored at the Sun \citep{Gosling_2006, Lavraud_2009, Lavraud_2020}.
This inward/outward asymmetry in both ion and electron signatures is reminiscent of the earthward/tailward energy flow asymmetry in magnetotail reconnection observed at the Earth \citep{Eastwood_2013, Tyler_2016, Lu_2018, Beyene_2024}. 
At the Sun, the slower inward jets could stem from the closed versus open in the field line topology on each side of the reconnection site, and be linked with the presence of an obstacle in the sunward flow \citep{Birn_Hesse_2005}.
The radial gradient in available magnetic energy, density, and temperature, as well as the unidirectional background flow speed, could also play a role in breaking the radial symmetry of HCS reconnection.

\subsection{Spectral properties of the turbulent cascade at the HCS}
\label{subsec: 7.3_turbulence}
\begin{figure}[t]
    \centering  \includegraphics[width=.49\textwidth]{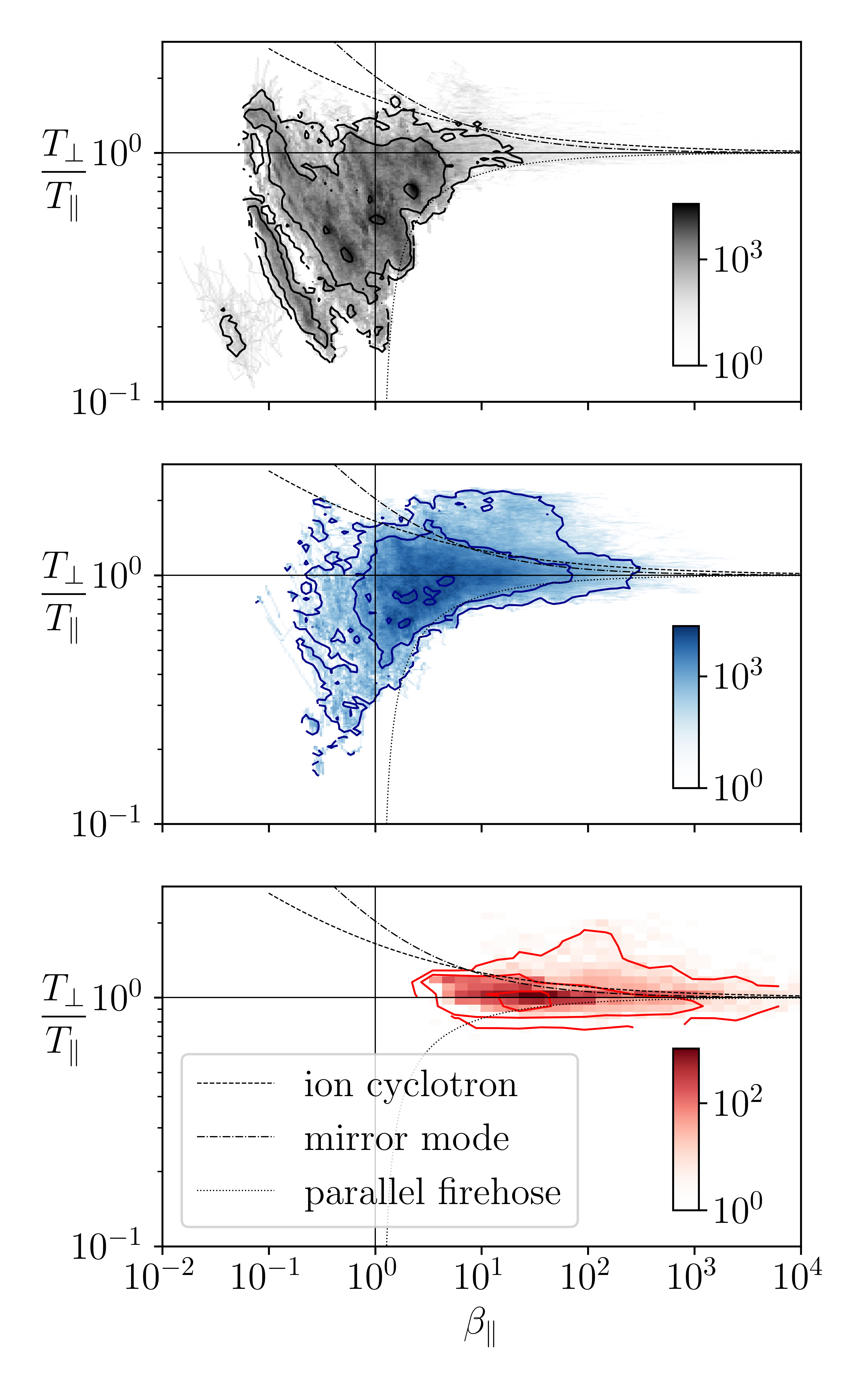}
    \caption{Distribution of bakground (a), HCS (b) and MH (c) plasma in the of the $\beta_{\parallel}$--$T_{\perp}/T_{\parallel}$ plane. Colorbars indicate the number of points in each bin.
    The background solar wind surrounding the HCS (a) was computed on time windows lasting 50\% of $\delta t$ on each side of HCS crossings (in order to get comparable count numbers).
    The MH plasma properties (c) were taken at the global minimum of each MH.
    Instability thresholds for the ion cyclotron, the mirror mode and the parallel firehose \citep{Verscharen_2019} are displayed in each panel.}
    \label{fig: brazil}
\end{figure}

We then study how the energy cascade behaves within and around the HCS. 
The magnetic field power spectra present a clean break around the ion gyro-frequency, with a -1.5 $\pm$ 0.1 spectral slope observed in the inertial range and a steeper slope below the ion scale.
These measurements are in agreement with previous observations of solar wind turbulence \citep[e.g., ][]{Alexandrova_2009, Bruno_2014, Sahraoui_2020, Chen_2020}.
We find that, for HCS crossings, the energy contained in the magnetic field fluctuations above the ion gyro-frequency is increased compared to the surrounding solar wind (Figure~\ref{fig: PSD}).
In that frequency range, the spectral slope of the magnetic field PSD inside HCS crossings reaches -3, while the surrounding HCS plasma has a steeper slope of -3.7 on average.
Given that magnetic reconnection is ubiquitous at the HCS, our finding is therefore particularly consistent with simulation work predicting that magnetic reconnection enhances the energy content of fluctuations below the ion scale \citep{Franci_2017, Cerri_Califano_2017, Franci_2018}, through the mean of sub ion-scale flux rope generation.
In particular, \citet{Franci_2017} show that the inverse cascade thus triggered combines with the forward cascade to form a PSD with a -3 spectral slope, which is consistent with our findings at the HCS.
The different spectral slopes could also be explained by a higher plasma $\beta$ inside the HCS compared to the background plasma, as shown in Figure~\ref{fig: brazil}. 
Indeed, observations \citep{Bruno_2014} and simulations \citep{Franci_2016} have shown that the slope of the magnetic field spectrum increases with $\beta$. 
In future work, it would be interesting to study the behavior of the energy transfer rate around and inside the HCS, and to check whether the turbulent cascade is stronger in the HCS.

In the inertial range, we distinguish three regions : the far-away surrounding solar wind (computed over a 150\%~$\delta t$ time window on each side), the close-by solar wind (computed over a 30\%~$\delta t$ time window on each side), and the HCS itself.
We note that, even with the farther away surrounding solar wind, we might still be capturing streamer belt plasma rather than coronal hole plasma.
This probably explains why both PSDs are quite close in the inertial range, while \cite{Chen_2021} report a higher energy content in the coronal hole solar wind compared to streamer belt plasma. 
We also retrieve the steeper magnetic spectrum of the HCS in the inertial range, observed by \citet{Chen_2021} during E4.

In parallel, we show that the energy content of the magnetic field fluctuations is lower near the HCS in the inertial range (Figure~\ref{fig: PSD}).
We link that decrease to a strong attenuation of large Alfvénic deflection occurrence -- i.e. magnetic switchback occurrence -- near the HCS, which we observe when browsing \textit{in situ} data.
This observation was previously reported for early PSP encounters \citep{Chen_2021, Woolley_2023_phd}, and our work is therefore a confirmation that quiet solar wind tends to surround the HCS.
This attenuation of large Alfvénic deflection occurrence may be due to a change in connectivity, as magnetic switchbacks are thought to originate from coronal hole winds \citep[e.g., ][]{Bale_2019, Badman_2020}. Close to the HCS, PSP is probably no longer connected to these coronal holes but is observing helmet streamer plasma instead, devoid of magnetic switchbacks. 

\subsection{Mirror mode storms in the HCS}
\label{subsec: 7.4_MH}
Finally, we report that linearly polarized MH trains are ubiquitous within HCS crossings close to the Sun by PSP (section \ref{sec: 6_MH_train}).
They are constituted of linear MHs with a size just above the ion gyro-radius (Figure~\ref{fig: MH_stats}), making them magneto-hydrodynamic structures. 
They can reach an occurrence of 16~min$^{-1}$ on average in the HCS, with a maximum occurrence rate of 54~min$^{-1}$ observed. 
These rates are much higher than previously reported occurrences of linear MHs observed by PSP in the inner heliosphere \citep{Yu_2021}, indicating that HCS crossings are a favored place to look for and study MHs.
In the literature, MH trains are also called mirror mode storms, as they are interpreted as remnants of mirror mode waves \citep{Russell_2009, EnriquezRivera_2013J, Dimmock_2022}, similar to what is observed in magnetospheric plasmas \citep{Volwerk_2008, Soucek_2008, Genot_2009}.
In recent work, \citet{Arro_2024} showed that large scale amplitude fluctuations in the solar wind evolve into stable MHs that can persist in the turbulent plasma. 

In Figure~\ref{fig: brazil}, we display the distribution of the background, HCS, and MH plasma in the $\beta_{\parallel}$--$T_{\perp}/T_{\parallel}$ plane, where $\beta_{\parallel} = n k_B T_{\parallel} \left(\dfrac{B^2}{2 \mu_0}\right) ^{-1}$, and $T_{\parallel}$ and $T_{\perp}$ are the ion temperatures parallel and perpendicular to the magnetic field.
Here, density and temperature have been linearly interpolated to the magnetic field measurements to compute $\beta_{\parallel}$.
The field-aligned temperatures were obtained by rotating the temperature tensor into the magnetic field aligned frame, and taking its diagonal components $T_{\parallel}$, $T_{\perp 1}$ and $T_{\perp 2}$.
We assumed perpendicular gyrotropy and took $T_{\perp}$ to be the maximum of $T_{\perp 1}$ and $T_{\perp 2}$, to reduce the impact of the solar wind being blocked by the heat shield in the SPAN-ion FOV.
As expected, we see that HCS plasma has a higher $\beta$ compared to the surrounding solar wind \citep{Huang_2023}.
It is also more isotropic, consistent with an energized isotropic core with no cold proton beam population, as has been previously reported within HCS outflows \citep{Lavraud_2020, Phan_2022}.
Anisotropies become increasingly prone to the mirror mode and the firehose instabilities as $\beta$ increases. 
A significant fraction of the HCS plasma seems to be mirror unstable.
These time intervals are directly associated with MH observations, lending support to the mirror mode origin of MHs. 
The unstable points beyond the mirror curve could hint at a still growing mirror instability, with unsaturated MHs.
We caution, however, that the lower time resolution on ion anisotropy could bias the measurements in that region.
In panel (c), we show the plasma properties measured at MH minima. 
It contains high $\beta$ plasma with anisotropy slightly over 1 on average (Q$_{1, 2, 3}$ = [0.99, 1.03, 1.06]).
The omnipresence of MH trains within HCS crossings shows that the mirror mode instability plays an important role in regulating the anisotropy of the plasma during HCS reconnection. 

\section{Summary and Conclusions}
In this manuscript, we report on properties of the near-Sun HCS observed by the PSP mission. We find that the HCS location can be stable across Carrington rotations, consistent with 1~AU results. 
The near-Sun HCS thickness is similar to 1~AU observations, with an average of 1.6$\times 10^5$~km.
The main difference in HCS properties lies in the occurrence of magnetic reconnection signatures,  and we find that 82\% (32/39) of near-Sun HCS crossings are reconnecting, while it is a rare observation at 1~AU.
We also report an increased occurrence of sunward reconnection jets closer to the Sun, and a difference in acceleration between sunward and anti-sunward jets. 
We find that turbulence levels are enhanced
in the ion kinetic range at the HCS, consistent with the triggering of an inverse cascade by magnetic reconnection.
Finally, we highlight the ubiquity of magnetic hole trains in the high $\beta$ environment of the near-Sun HCS, a novel observation compared to 1~AU HCS crossings.

Our findings shed new light on the properties of magnetic reconnection in the high $\beta$ plasma environment of the HCS, its interplay with the turbulent cascade, and the importance of the mirror mode instability in regulating the temperature anisotropy.
The HCS is a key structure to study the fundamental processes of plasma physics in space. 
Future work could focus on energy balance at the HCS, to better understand energy conversion associated with large scale magnetic reconnection close to our star. 
Understanding the temporal evolution of HCS dynamics and helmet streamers could also be pursued through the detailed study of E17, both \textit{in situ} and remotely. 
Finally, conjunction studies with the Solar Orbiter spacecraft and 1~AU space missions would unveil how HCS reconnection jets evolve and dissipate in the heliosphere.

\begin{acknowledgments}

We acknowledge the NASA Parker Solar Probe Mission and particularly the FIELDS team led by S. D. Bale and the SWEAP team led by J. Kasper for the use of data. Parker Solar Probe was designed, built, and is now operated by the Johns Hopkins Applied Physics Laboratory as part of NASA’s Living with a Star (LWS) program (contract NNN06AA01C). The data used in this study are available at the NASA Space Physics Data Facility (SPDF): \href{https://spdf.gsfc.nasa.gov}{https://spdf.gsfc.nasa.gov}, and at the PSP science gateway \href{https://sppgway.jhuapl.edu/}{https://sppgway.jhuapl.edu/}.     
We visualize data using both the \href{http://amda.cdpp.eu/}{AMDA} science analysis system and the \href{https://speasy.readthedocs.io/en/latest/index.html}{speasy} python package  provided by the Centre de Données de la Physique des Plasmas (CDPP) supported by CNRS, CNES, Observatoire de Paris, and Université Paul Sabatier (UPS), Toulouse. 
Work by NF was supported by CNES, CNRS, and the UKRI/STFC grant ST/W001071/1. 
JPE acknowledges support from UKRI/STFC grant ST/W001071/1.
TDP acknowledges support from NASA grants 80NSSC20K1781 and 80NSSC25K7678.
L.F. is supported by the Royal Society University Research Fellowship No. URF/R1/231710.
This research was supported by the International Space Science Institute (ISSI) in Bern, through the ISSI International Team project 24-612: Excitation and Dissipation of Kinetic-Scale Fluctuations in Space Plasmas.
\end{acknowledgments}

\appendix

\section{Timetables of events}
 
\bottomcaption{Timetable of 39 HCS crossings identified by PSP during E6-E21 while below $50~\mathrm{R_{\odot}}$. Colums respectively indicate the event number, its associated encounter, the year and month of occurrence, start and end time of the HCS in the form ddThh:mm:ss. The last column indicate the presence of and outward (+) or inward (-) jet. Asterisk refer to events where a smaller jet (Table~\ref{tab: app_jets}) was also detected at $B_R$ reversal}
\label{tab: app_HCS}
\begin{supertabular}{cccccc}
\# & E$_x$ & yyyy-mm & $t_i$ & $t_f$ & jet \\
\hline 
1	&	E6	&	2020-09	&	25T17:45:00	&	25T19:21:00	&	~*	\\
2	&	E7	&	2021-01	&	17T13:14:00	&	17T13:31:00	&	+	\\
3	&	E7	&	2021-01	&	19T21:09:00	&	19T23:16:00	&	+	\\
4	&	E8	&	2021-04	&	24T16:02:30	&	24T16:03:10	&		\\
5	&	E8	&	2021-04	&	29T00:54:15	&	29T01:54:00	&	+*	\\
6	&	E8	&	2021-04	&	29T08:14:00	&	29T08:29:00	&	+	\\
7	&	E8	&	2021-04	&	29T13:40:30	&	29T14:24:00	&	~*	\\
8	&	E9	&	2021-08	&	10T00:28:00	&	10T01:53:00	&	+	\\
9	&	E9	&	2021-08	&	10T10:35:00	&	10T10:42:00	&	+	\\
10	&	E9	&	2021-08	&	10T11:04:00	&	10T11:14:00	&		\\
11	&	E9	&	2021-08	&	10T17:57:00	&	10T18:51:00	&		\\
12	&	E11	&	2022-02	&	25T12:28:00	&	25T12:36:00	&	+	\\
13	&	E12	&	2022-06	&	02T17:24:00	&	02T17:36:00	&	+	\\
14	&	E13	&	2022-09	&	06T17:27:00	&	06T17:40:00	&	+	\\
15	&	E14	&	2022-12	&	12T06:18:00	&	12T10:00:00	&	-	\\
16	&	E15	&	2023-03	&	16T04:40:00	&	16T04:44:00	&	+	\\
17	&	E15	&	2023-03	&	17T21:03:00	&	17T21:14:00	&		\\
18	&	E16	&	2023-06	&	22T00:23:00	&	22T01:54:00	&	+	\\
19	&	E16	&	2023-06	&	22T04:40:00	&	22T06:08:00	&	+*	\\
20	&	E16	&	2023-06	&	24T04:44:00	&	24T08:00:00	&	+	\\
21	&	E17	&	2023-09	&	25T20:44:00	&	25T23:32:00	&	+	\\
22	&	E17	&	2023-09	&	26T16:47:00	&	26T17:24:00	&		\\
23	&	E17	&	2023-09	&	26T18:59:00	&	26T19:08:00	&		\\
24	&	E17	&	2023-09	&	26T20:18:00	&	26T21:31:00	&	+	\\
25	&	E17	&	2023-09	&	27T19:50:00	&	27T20:00:00	&	-	\\
26	&	E17	&	2023-09	&	28T06:18:00	&	28T07:03:00	&	+	\\
27	&	E18	&	2023-12	&	29T02:21:00	&	29T02:53:00	&	-	\\
28	&	E19	&	2024-03	&	29T23:08:00	&	29T23:15:00	&	+	\\
29	&	E19	&	2024-03	&	30T12:34:00	&	30T12:56:00	&	-	\\
30	&	E20	&	2024-06	&	29T12:28:00	&	29T12:35:00	&	+	\\
31	&	E20	&	2024-06	&	29T23:51:00	&	30T00:42:00	&	-	\\
32	&	E20	&	2024-07	&	01T17:07:00	&	01T18:33:00	&	+	\\
33	&	E20	&	2024-07	&	04T05:01:00	&	04T05:22:00	&		\\
34	&	E21	&	2024-09	&	26T05:47:00	&	26T10:39:00	&	+	\\
35	&	E21	&	2024-09	&	29T13:05:00	&	29T13:18:00	&	-	\\
36	&	E21	&	2024-09	&	30T01:55:19	&	30T01:55:54	&	-	\\
37	&	E21	&	2024-09	&	30T11:51:00	&	30T12:39:00	&	+	\\
38	&	E21	&	2024-10	&	04T03:22:00	&	04T03:47:00	&	+	\\
39	&	E21	&	2024-10	&	04T06:31:00	&	04T06:46:00	&	+	\\
\hline
\hline
\end{supertabular}

\bottomcaption{Timetable of HCS partial crossings identified by PSP during E6-E21 while below $50~\mathrm{R_{\odot}}$}
\label{tab: app_partials}
\begin{supertabular}{ccccc}
\# & E$_x$ & yyyy-mm & $t_i$ & $t_f$  \\
\hline 
1	&	E6	&	2020-09	&	25T08:43:00	&	25T14:20:00	\\
2	&	E6	&	2020-09	&	30T08:40:00	&	30T17:20:00	\\
3	&	E7	&	2021-01	&	17T13:42:00	&	17T14:42:00	\\
4	&	E7	&	2021-01	&	19T13:15:00	&	19T17:20:00	\\
5	&	E8	&	2021-04	&	29T09:23:00	&	29T10:23:00	\\
6	&	E9	&	2021-08	&	10T13:30:00	&	10T17:00:00	\\
7	&	E10	&	2021-11	&	22T01:10:00	&	22T01:50:00	\\
8	&	E10	&	2021-11	&	22T02:10:00	&	22T02:38:00	\\
9	&	E12	&	2022-06	&	02T14:00:00	&	02T14:30:00	\\
10	&	E12	&	2022-06	&	02T19:28:00	&	02T19:51:00	\\
11	&	E15	&	2023-03	&	14T22:47:00	&	15T11:15:00	\\
12	&	E16	&	2023-06	&	24T08:30:00	&	24T09:50:00	\\
13	&	E16	&	2023-06	&	25T05:15:00	&	25T11:17:00	\\
14	&	E17	&	2023-09	&	23T16:30:00	&	23T17:50:00	\\
15	&	E17	&	2023-09	&	23T20:00:00	&	25T01:17:00	\\
16	&	E17	&	2023-09	&	25T15:53:00	&	25T17:44:00	\\
17	&	E17	&	2023-09	&	25T18:40:00	&	25T19:25:00	\\
18	&	E17	&	2023-09	&	25T19:42:00	&	25T20:07:00	\\
19	&	E17	&	2023-09	&	26T00:42:00	&	26T01:16:00	\\
20	&	E17	&	2023-09	&	26T05:52:00	&	26T08:32:00	\\
21	&	E17	&	2023-09	&	26T12:00:00	&	26T14:10:00	\\
22	&	E20	&	2024-06	&	29T12:16:00	&	29T12:19:00	\\
23	&	E20	&	2024-07	&	01T20:30:00	&	02T02:14:00	\\
24	&	E20	&	2024-07	&	03T12:30:00	&	03T15:00:00	\\
25	&	E21	&	2024-09	&	29T09:50:00	&	29T10:00:00	\\
26	&	E21	&	2024-09	&	29T10:30:00	&	29T11:00:00	\\
27	&	E21	&	2024-09	&	30T01:56:40	&	30T01:57:33	\\
28	&	E21	&	2024-09	&	30T02:19:00	&	30T02:31:00	\\
29	&	E21	&	2024-10	&	04T06:17:00	&	04T06:24:00	\\
\hline
\hline
\end{supertabular}

\bottomcaption{Timetable of small ion jets located at $B_R$ reversals on the side of an HCS crossing (see section \ref{subsec: 4_Rx_id} for more details)}
\label{tab: app_jets}
\begin{supertabular}{cccccc}
\# & E$_x$ & yyyy-mm & $t_i$ & $t_f$ & jet \\
\hline 
1	&	E06	&	2020-09	&	25T17:45:00	&	25T17:51:00	& +\\
2	&	E08	&	2021-04	&	29T00:54:15	&	29T00:56:00	& +\\
3	&	E08	&	2021-04	&	29T13:40:50	&	29T13:42:00	& - \\
4	&	E16	&	2023-06	&	22T04:41:36	&	22T04:42:10	& - \\
\hline
\hline
\end{supertabular}

\section{Parameter variation inside the HCS}
\label{sec: app_vl}

\begin{figure}[h!]    
    \includegraphics[width=.48\textwidth]{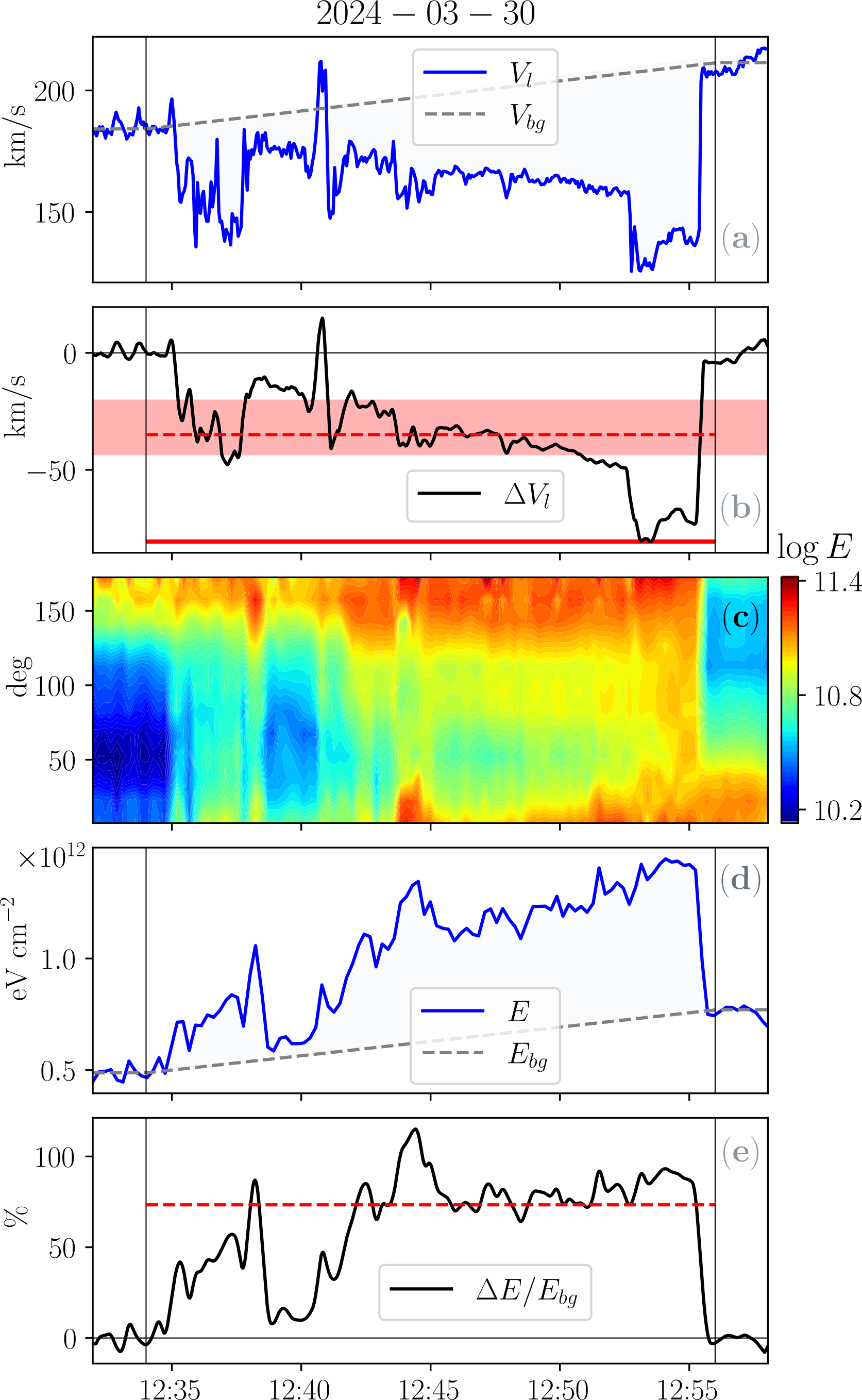}
    \caption{
    Illustration of the computation of $\Delta V_l$ and $\Delta E$. 
    We display HCS crossing \#29 delimited by vertical black lines and surrounded by time windows lasting 10\% of its duration. 
    We show (a) $V_l$ and $V_{bg}$; (b) the smoothed $\Delta V_l$; (c) electron PAD in the energy range 300--800~eV; (d) the energy flux in the same energy range; and (e) the relative variation of this energy flux.
    In panels (b) and (e), median values are highlighted as a horizontal red dashed line. 
    In panel (b), the red shaded area shows the Q$_1$ -- Q$_3$ quartiles of  $\Delta V_l$ over the HCS crossing, and the value of maximum $V_l$ variation is indicated as a horizontal red line. 
    }
    \label{fig: app_dV_dE}
\end{figure}

We here detail how we get the velocity variations displayed in Figure~\ref{fig: dV_Va} and the strahl energy flux relative variation shown in Figure~\ref{fig: dE}.
We use event \#29 of Table~\ref{tab: app_HCS} as an illustration.

For each event, we rotate the velocity in its associated $lmn$ frame and focus on the $l$ component of the solar wind velocity.
A background velocity $V_{bg}$ is defined as the linear interpolation between median values of $V_l$ on each side of the HCS crossing, computed over time windows lasting 10\% of the crossing duration.
We then get $\Delta V_l = V_l - V_{bg}$ and smooth it to remove noisy variations. 
In panels (a) and (b) of Figure~\ref{fig: app_dV_dE}, we show these different quantities for event \#29.
This particular HCS crossing is occurring during E19 and is located at 13~\Rs.
The hybrid Alfvén speed is 165~\kms while the quartiles Q$_{1, 2, 3}$ of $\Delta V_l$ are respectively [-43, -35, -20]~\kms, and its maximum speed variation is -81~\kms.
This yields $\mathcal{Q}_2 (\Delta V_l) / V_A= -0.21^{+0.09}_{-0.05}$ and $\max(\Delta V_l)/V_A = -0.49$. 

We apply a similar approach to estimate the strahl energy flux variation, with the background energy flux estimated using the same linear interpolation.
In panel (c), we clearly see the bidirectional signature in the strahl occurring during the inward reconnection jet.
The associated relative energy flux increase $(E - E_{bg})/E_{bg}$ reaches a median value of 73\%.

\section{Magnetic hole detection}
\label{sec: app_mh}

\begin{figure}[h!]    
    \includegraphics[width=.48\textwidth]{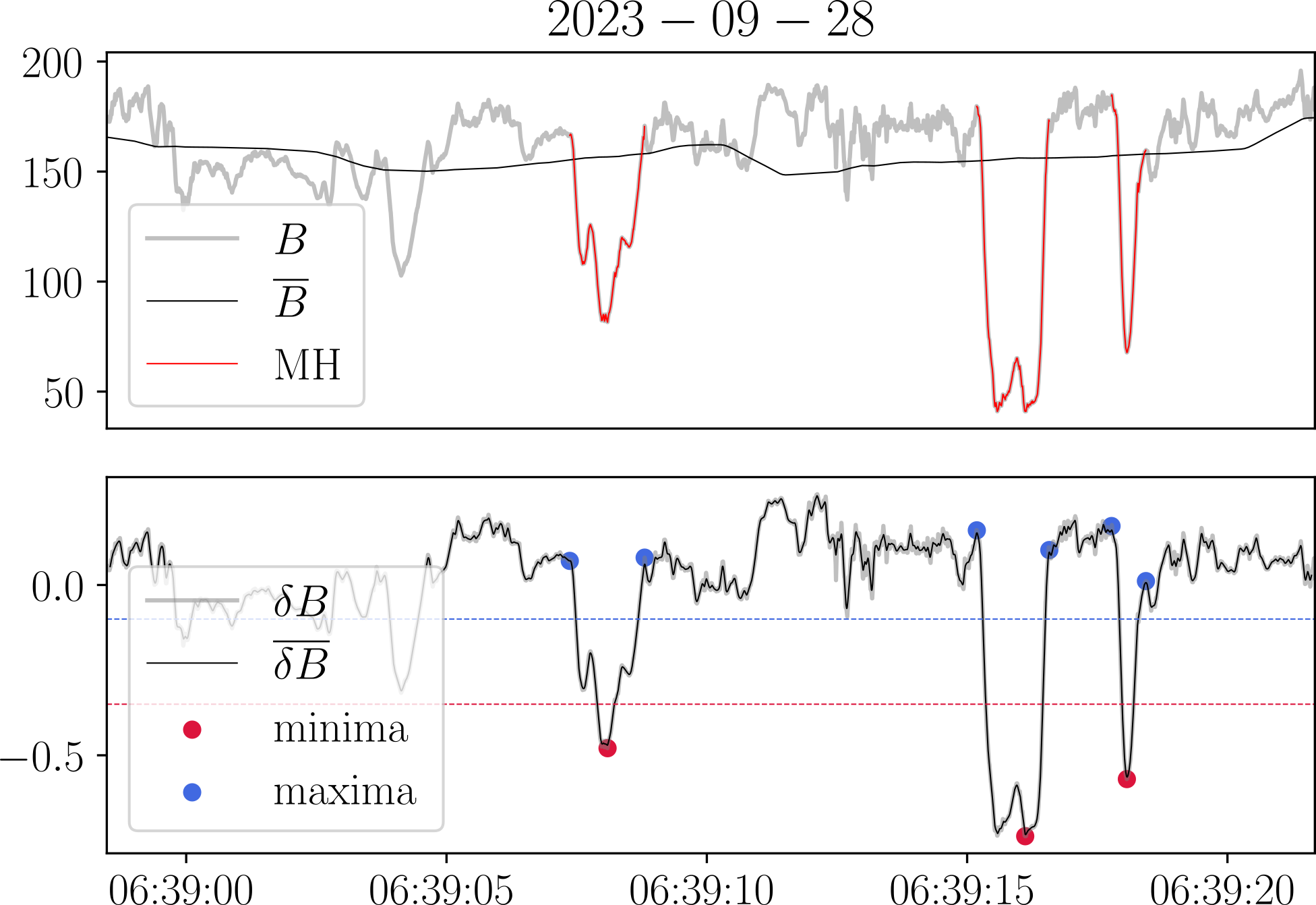}
    \caption{
    Illustration of MH identification during event \#26. The top panel shows the magnetic field amplitude $B$ and its 10~s average $\overline{B}$. 
    Detected MHs are highlighted in red.
    The bottom panel shows the relative variation  $\delta B$ and its smoothed version $\overline{\delta B}$. Red dots show the global minimum of each MH, blue dots are the closest maxima identified as the MH limits. Horizontal lines indicate the thresholds used for minima (-0.35, red) and maxima (-0.1, blue).
    }
    \label{fig: app_MH}
\end{figure}
In this section, we describe how we automatically detect magnetic holes \textit{in situ}.
We implement a hysteresis approach: a magnetic hole is detected if the relative variation of the magnetic field amplitude is lower than -35\%. 
The MH limits are then defined as the closest maxima where the relative variation is more than -10\%.
In practice, we implement the following steps:
\begin{enumerate}    
    \item we compute $\overline{B}$, a 10~s average of $B$;
    \item we compute $\delta B = (B - \overline{B}) /  \overline{B}$ the relative variation of the magnetic field amplitude;
    \item we compute $\overline{\delta B}$, which is $\delta B$ smoothed with a 0.05~s Gaussian filter to remove noisy variations;
    \item we find all minima of $\overline{\delta B}$ where $\delta B<-0.35$ within the time period;
    \item we find all maxima of $\overline{\delta B}$ where $\delta B>-0.1$ within the time period;
    \item for each minima detected in (4), we flag as an MH the interval between the closest maxima identified in (5);
    \item for each detected MH, we identify the location of the global minimum.
\end{enumerate}
In Figure~\ref{fig: app_MH}, we show these quantities and the identification results for three MHs observed during event \#26.


\bibliography{sample701}{}
\bibliographystyle{aasjournalv7}



\end{document}